\documentclass[10pt, conference, letterpaper]{IEEEtran}
\IEEEoverridecommandlockouts
\usepackage{cite}
\usepackage{amsmath,amssymb,amsfonts}
\usepackage{algorithmic}
\usepackage{graphicx}
\usepackage{textcomp}
\usepackage{xcolor}
\usepackage{subfigure}
\usepackage{tikz}
\usepackage{algorithm}  
\usepackage{xspace}
\usepackage{url}
\usepackage{comment}
\usepackage{caption}
\usepackage{diagbox}
\usepackage{balance}
\usepackage{threeparttable}
\usepackage{multirow}

\def\BibTeX{{\rm B\kern-.05em{\sc i\kern-.025em b}\kern-.08em
    T\kern-.1667em\lower.7ex\hbox{E}\kern-.125emX}}
\RequirePackage[normalem]{ulem} 
\RequirePackage{color}\definecolor{RED}{rgb}{1,0,0}\definecolor{blue}{rgb}{0,0,1} 
    
\renewcommand{\baselinestretch}{1.1}
\definecolor{revision}{RGB}{0,0,255}

\newcommand{\revstart}{\begin{color}{revision}}
\newcommand{\revend}{~\!\!\end{color}}
\newcommand{\pname}{\texttt{AccEar}\xspace}

\title{AccEar: Accelerometer Acoustic Eavesdropping with Unconstrained Vocabulary}
\author{
    \IEEEauthorblockN{
        Pengfei Hu\IEEEauthorrefmark{1},
        Hui Zhuang\IEEEauthorrefmark{1},
        Panneer Selvam Santhalingam\IEEEauthorrefmark{2},
        Riccardo Spolaor\IEEEauthorrefmark{1},
        Parth Pathak\IEEEauthorrefmark{2}, \\
        Guoming Zhang\IEEEauthorrefmark{1},
        Xiuzhen Cheng\IEEEauthorrefmark{1}
    }
    \IEEEauthorblockA{
        \IEEEauthorrefmark{1}
        Shandong University, China\\
        \IEEEauthorrefmark{2}
        George Mason University, USA\\
        Email: \{phu, rspolaor, guomingzhang, xzcheng\}@sdu.edu.cn, \{psanthal, phpathak\}@gmu.edu, \{zhuanghui303\}@gmail.com
    }
}
\begin{document}
\maketitle
\begin{abstract}
With the increasing popularity of voice-based applications, acoustic eavesdropping has become a serious threat to users' privacy. 
While on smartphones the access to microphones needs an explicit user permission, 
acoustic eavesdropping attacks can rely on motion sensors (such as accelerometer and gyroscope), which access is unrestricted. 
However, previous instances of such attacks can only recognize a limited set of pre-trained words or phrases. 
In this paper, we present \pname, an accelerometer-based acoustic eavesdropping attack 
that can reconstruct any audio played on the smartphone's loudspeaker with unconstrained vocabulary. 
We show that 
an attacker can employ a conditional Generative Adversarial Network (cGAN) to reconstruct high-fidelity audio from low-frequency accelerometer signals. 
The presented cGAN model learns to recreate high-frequency components of the user's voice from low-frequency accelerometer signals through spectrogram enhancement. 
We assess the feasibility and effectiveness of \pname attack in a thorough set of experiments using audio from 16 public personalities. 
As shown by the results in both objective and subjective evaluations, \pname successfully reconstructs user speeches from accelerometer signals in different scenarios including varying sampling rate, audio volume, device model, etc. 

\end{abstract}
\section{INTRODUCTION}




Nowadays, voice-based applications (e.g., voice over IP, video conferencing, voice assistants) on smartphones are part of our daily lives. 
Since the audio from such applications can reveal private information about the user, mobile operating systems grant access to the microphone only with explicit user permission. 
To bypass this restriction, security researchers leverage the unrestricted motion sensors  (e.g., accelerometer, gyroscope) as a side-channel to carry out acoustic eavesdropping attacks\cite{michalevsky2014gyrophone, zhang2015accelword, anand2018speechless, ba2020learning,han2017pitchln}. 
These side-channel attacks are possible since motion sensors are sensitive to the vibrations produced by sound waves. 
From motion sensors data, these prior works can recognize words/phrases that are either spoken by the user or emitted from the smartphone's speaker.


While effective, most of prior attacks of audio eavesdropping using motion sensors treat the audio extraction problem as a classification problem. Here, an attacker can create signatures of motion sensor data for different words or phrases and can recognize them using a machine learning model. However, such an attack is primarily limited to the pre-trained set of words and phrases and does not work well in reconstructing any unknown audio signals. Ba \textit{et al.}~\cite{ba2020learning} propose a deep neural network based approach for speech reconstruction, however they can only recover the partial vowels in low frequency region (below $1500$Hz). The low sampling rate of motion sensors imposes a limit, making the complete reconstruction of audio an extremely challenging problem.

In this work, we present \pname, a new type of accelerometer-based eavesdropping attack that can reconstruct any audio signal with unconstrained vocabulary. It uses the accelerometer signals measured on a smartphone while the audio is being played on the built-in smartphone speaker. Given that the sampling rate of the accelerometer is limited (maximum of $500$Hz) by the mobile operating systems, the low-frequency, low-resolution signal cannot be directly used for audio reconstruction. We address this challenge by developing Conditional Generative Adversarial network (cGAN)\cite{2014Conditional} based model that infers and recreates the high frequency components based on the measured low-frequency accelerometer signal. Through a limited amount of training set, our cGAN-based model can learn the mapping between low-frequency accelerometer data and the corresponding phonemes that they represent, enabling us to reconstruct any audio signal (e.g., words, phrases, sentences, etc.) that is unknown to the model (not used in training). For achieving this reconstruction, we design our cGAN model to operate on spectrograms where it learns to generate the complete audio spectrogram from the given low-frequency accelerometer signal spectrogram. The generated enhanced spectrograms are then used along with the Griffin-Lim algorithm\cite{griffin1984signal} to \emph{reconstruct clear, human-perceivable audio}.

Since our presented attack is not limited to the specific pre-trained set of words or phrases, it greatly increases the risk of information leakage in a wide range of commonly occurring scenarios. Some of the scenarios are listed below:
\begin{itemize}
    \item When a remote contact talks, shares videos or sends voice messages to a user via smartphone, an attacker can reconstruct the remote contact's voice to steal private information using \pname.
    \item An attacker can listen to user's voice memos or commands that may contain confidential information such as passwords, schedules, phone numbers, social security numbers, passcodes, etc. 
    \item When the user uses voice navigation, the attacker can use \pname to infer user's location and other preferences such as the type of location user likes to visit, restaurants, points-of-interest, etc.
    \item When the user's smartphone plays an audio that may contain a specific product name, the attacker can learn about the user's preferences of products, medical conditions, etc.
    \item The attacker can intercept the (voice-based) verification codes commonly used in two-factor authentications to obtain the access to user's account. 
\end{itemize}



\par Our contributions can be summarized as follows:
\begin{enumerate}
     \item We propose \pname, an acoustic eavesdropping system that uses accelerometer data to accurately reconstruct the user speech played by the smartphone speaker. 
    To the best of our knowledge, \pname is the first method that actually recovers the speech content with an unconstrained vocabulary rather than recognizing individual hot words/phrases.
    
    
    \item Our proposed method converts low-frequency accelerometer data into a comprehensible
    audio signal.
    To do so, we train cGAN models to learn the mapping between accelerometer data and the correspondent audio played by the smartphone speaker.
    The cGAN model can enrich an accelerometer signal by adding its missing high-frequency components and using the previously learned mapping to produce an audio signal.
    Our method demonstrates that cGAN can substantially enhance an attacker's capabilities even when the available data has limited resolution due to hardware or software restrictions.
    
    \item We carry out an extensive evaluation of \pname attack using an audio dataset from $16$ public personalities and several real-world scenarios. 
    \pname achieves an average Mel-Cepstral Distortion (a lower value indicates a better reconstruction performance) of $4.784$, a Mean Opinion Score (a higher value indicates a better reconstruction performance) of $3.637$, and an average Word Error Rate (a lower value indicates a better reconstruction performance) of $13.434\%$ for twenty volunteers.
    Through cross-user training, we also demonstrate that \pname can effectively reconstruct audio even when no audio samples of the victim are available for the training.
    
\end{enumerate}


The remaining paper is organized as follows. 
Section~\ref{sec:related} discusses the related work. 
Section~\ref{sec:preliminary} discusses the preliminaries of accelerometer, phoneme, and GAN. 
In Section~\ref{sec:system}, we present our system and describe its components in detail.
Section~\ref{sec:evalution} performs the evaluation on our system.
In Section~\ref{sec:discussion}, we discuss the obtained results, meaningful insights, and limitations of our work.
Section~\ref{sec:conclusion} summaries our work.







\section{Related Work}
\label{sec:related}
In this section, we introduce the works related to speech reconstruction via IMU (Inertial Measurement Unit) and other acoustic eavesdropping methods.
\subsection{Acoustic eavesdropping attacks via IMU}
In recent years, some security researchers focus on eavesdropping via motion sensors in smartphones as the motion sensors are sensitive and precise enough to capture the vibrations emitted by the object.
\par Michalevsky \textit{et al.}~\cite{michalevsky2014gyrophone} show that the gyroscopes in smartphones are sufficiently sensitive to measure acoustic signals in their vicinity. 
The authors place a smartphone and an active loudspeaker (i.e., playing sound) on the same solid surface. 
The sound emitted by the loudspeaker passes through the solid surface, which vibrations influence the readings of the smartphone's built-in gyroscope. 
Through analyzing the gyroscope measurements, they enable to recognize the person's identity and even retrieve some particular speech information. 
However, IMU data can only preserve information from frequencies below 200Hz, which results in a low accuracy ($77\%$) of digits recognition.

\begin{figure*}[t]
    \centering
    \subfigure[The accelerometer data spectrogram of vowels]{
    \includegraphics[width=0.47\columnwidth]{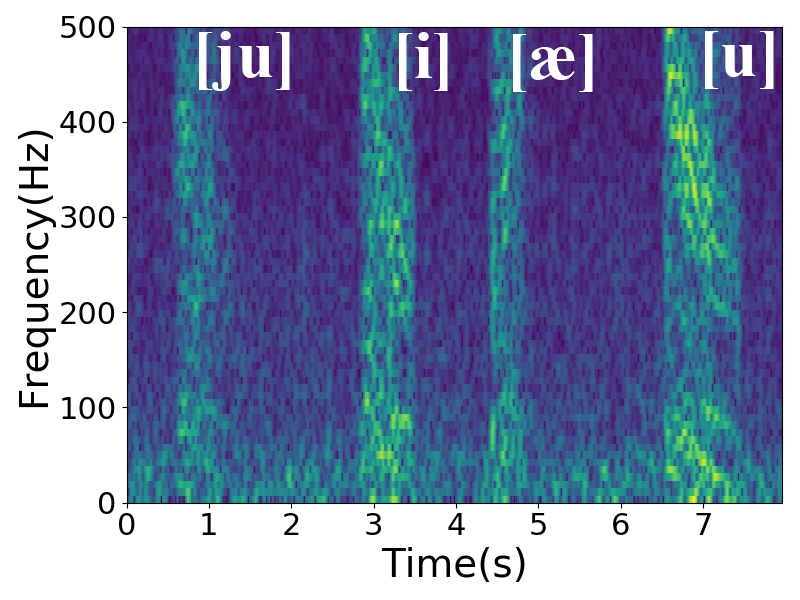}
    \label{fig:vowel_new}}
    \subfigure[The spectrogram of vowels]{
    \hspace{-12pt}\includegraphics[width=0.47\columnwidth]{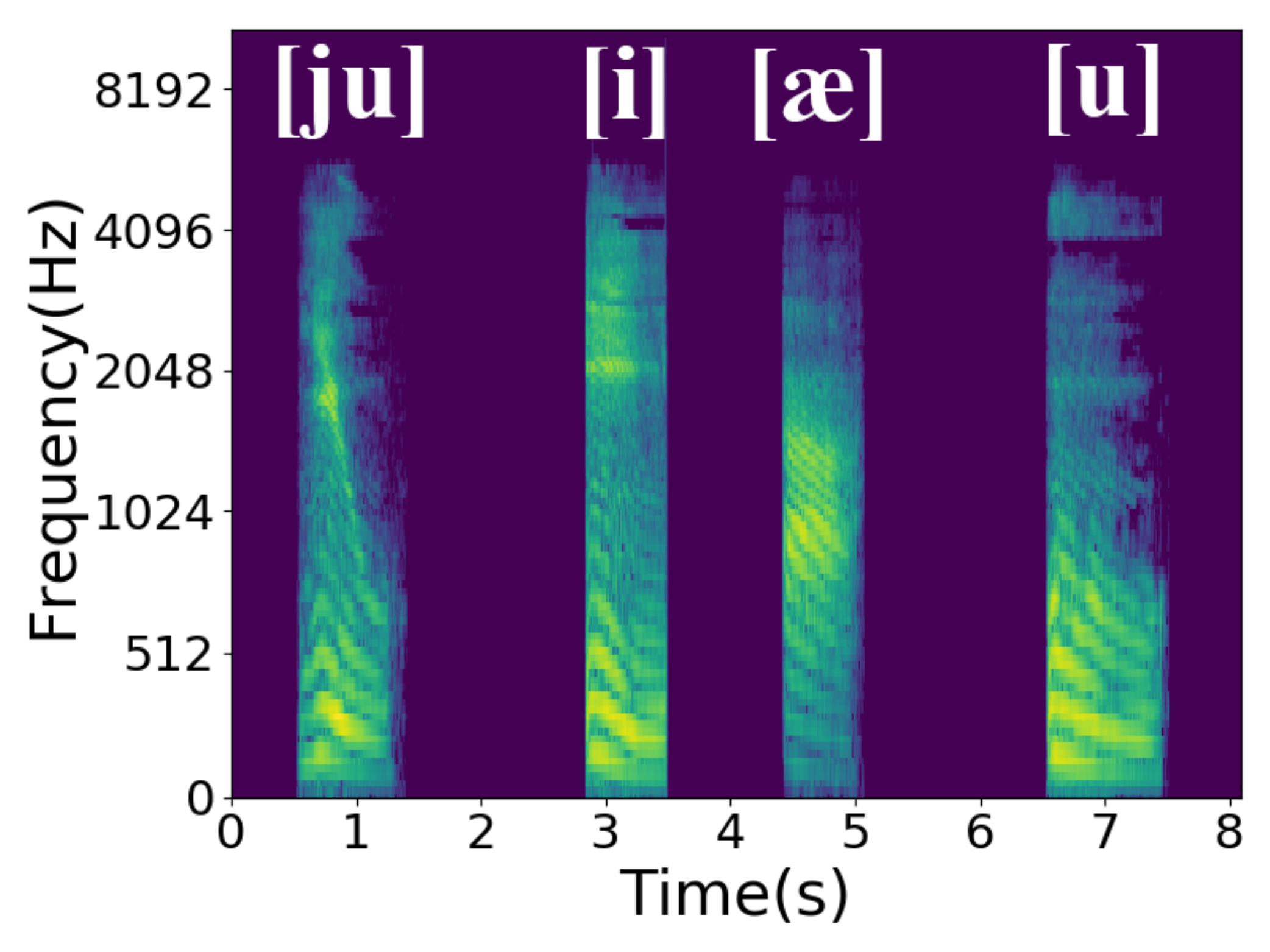}
    \label{fig:vowel_new_mel}
    }
    \subfigure[The accelerometer data spectrogram of consonants]{
    \includegraphics[width=0.47\columnwidth]{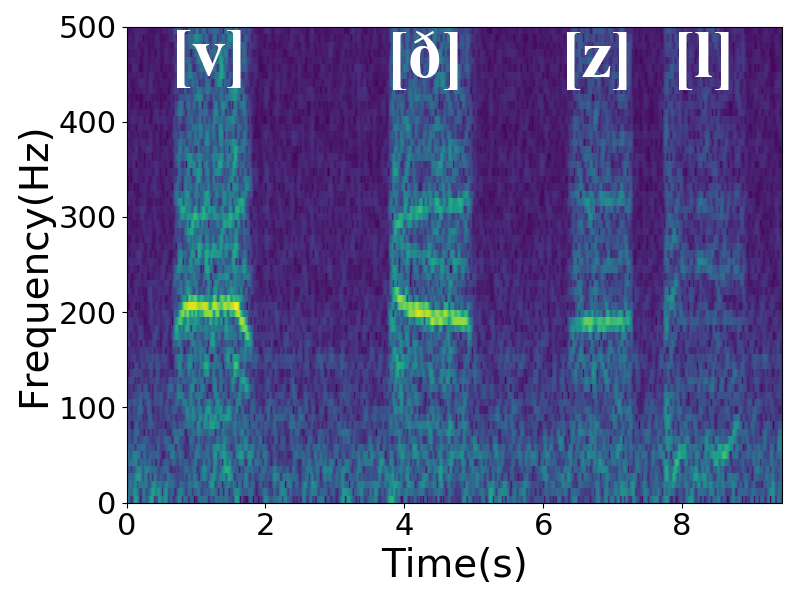}
    \label{fig:consonant_new}}
    \subfigure[The spectrogram of consonants]{
    \hspace{-12pt}\includegraphics[width=0.47\columnwidth]{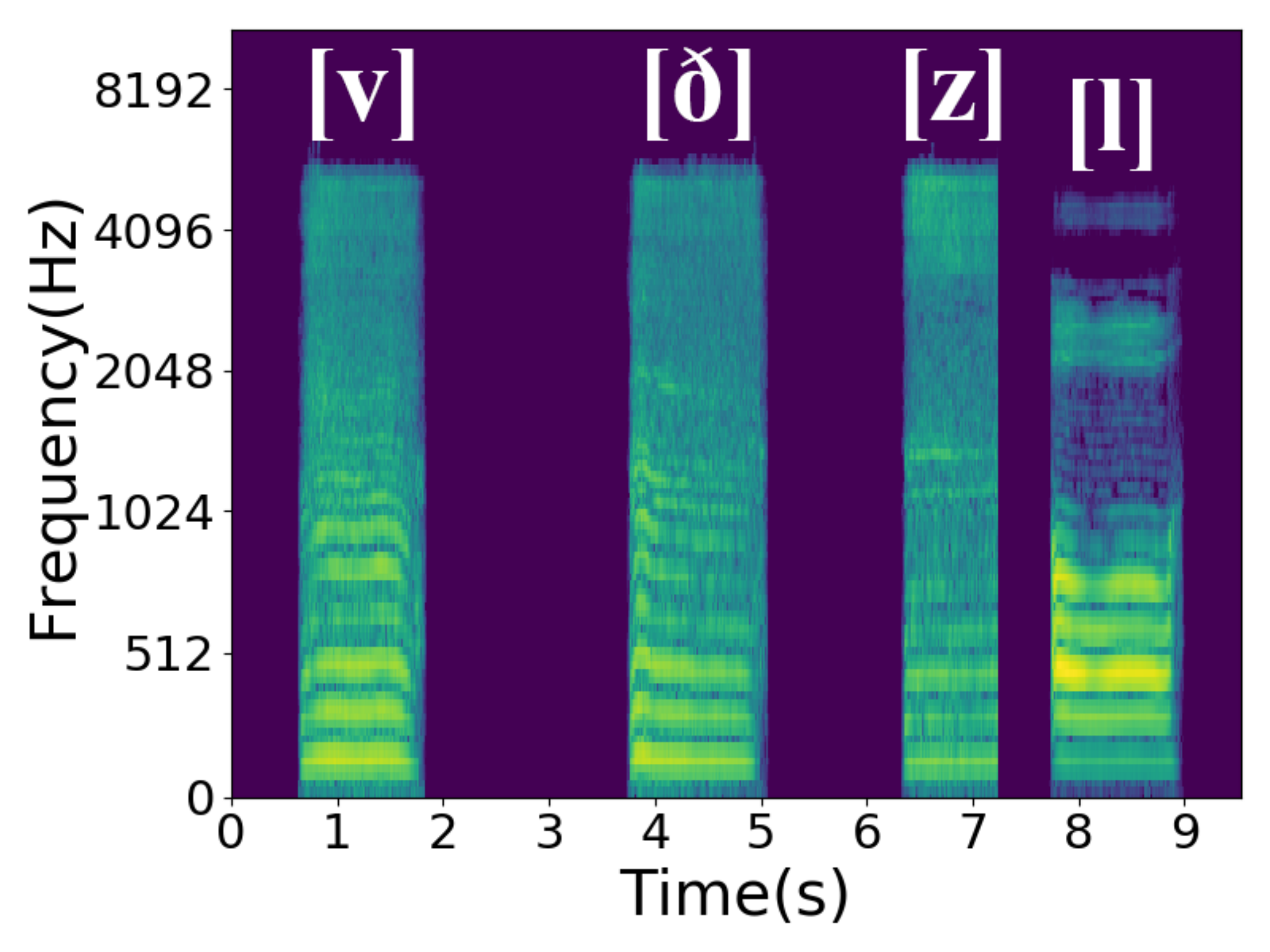}
    \label{fig:consonant_new_mel}
    }
    \caption{Spectrogram of phonemes }
    \label{fig:phoneme}
\end{figure*}

Zhang \textit{et al.}~\cite{zhang2015accelword} assess that accelerometers are also sensitive to the human voice. 
The authors hold the smartphone in their hands or place it on the desk and speak to the phone, which will cause the vibration of the accelerometer. Through observing the changes in the accelerometer data, they observe the vibration has specific pattern related to human's spoken words, and it is possible to extract the unique signatures of the hot words from the accelerometer data. Based on this observation, they design AccelWord to recognize the hot words such as ``Okay Google" or ``Hi Galaxy" from accelerometer data. 
However, Anand \textit{et al.}~\cite{anand2018speechless} argue that both human- and machine-rendered speech is not powerful enough to affect smartphone motion sensors through the air. 

More recently, Ba \textit{et al.}~\cite{ba2020learning} propose a new side-channel attack which eavesdrops on the speaker based on the accelerometer on the same smartphone. The vibration produced by the speaker can propagate through the motherboard and induce strong response on the accelerometer\cite{anand2021spearphone,ba2020learning}. Hence they can utilize the accelerometer measurements to recognize the sensitive information speech emitted by the speaker. They employ a deep neural network to further improve hot words recognition, which could achieve an accuracy of $99\%$ for digits only and $87\%$ for the combination of digits and letters. However, this deep neural network fails to reconstruct phonemes in high frequency (above $1500$Hz), which renders it incapable to perform full speech reconstruction.

\par All aforementioned works share the same disadvantage that they can only recognize or reconstruct hot words from the pre-established vocabulary. 
Since audio emitted by the speaker in a real-world scenario typically carries much more information instead of hot words solely,
such a limitation drastically reduces the amount of speech privacy that can be inferred.

\subsection{Other acoustic eavesdropping attacks}
Nowadays, the works related to eavesdropping have been extensively studied. Davis \textit{et al.}\cite{davis2014visual} recover sounds from high-speed footage of a variety of objects with different properties, such as a glass of water or a bag of chips, by using the principle that sound hitting an object causes the surface of the object to vibrate sightly. Kwong \textit{et al.}\cite{kwong2019hard} demonstrate that the mechanical components in magnetic hard disk drives are sensitive enough to extract and parse human speech. Guri \textit{et al.}\cite{guri2017speake} introduce the malware ``SPEAKE(a)R", which enables to turn the headphones, earphones, or earbuds connected to a personal computer into microphones when the standard microphone is not working or tapped. 
Roy \textit{et al.}\cite{roy2016listening} demonstrate that the vibration motor in mobile devices enables them to serve as a microphone by processing their response to the air vibrations from nearby sounds. Wang \textit{et al.}\cite{wang2016we} access the information of human conversations by detecting and analyzing the fine-grained radio reflections from mouth movements. Wei \textit{et al.}\cite{wei2015acoustic} use the acoustic-radio transformation (ART) algorithm to recover the sounds of the speaker device.
Muscatell \textit{et al.}\cite{muscatell1984laser} use a laser transceiver to eavesdrop on the sound in the room. In particular, the authors use a laser generator to shoot a laser onto an object in the room and a laser receiver to receive the reflected laser back. They can recover the sound by analyzing the reflected laser. 
Nassi \textit{et al.}\cite{nassi2020lamphone} use the hanging bulb and remote electro-optical sensor to eavesdrop sounds. The authors show that the sound causes the air pressure on the surface of the bulb to fluctuate so that the lamp is slightly vibrated. Then they use the electro-optical sensor to analyze the hanging bulb's frequency response to sound to recover the sound. 
\section{Preliminaries}
\label{sec:preliminary}

In this section, we briefly introduce the principles of the accelerometer, the characteristics of phonemes, and the idea of generative adversarial networks. We also provide references for an in-depth understanding of those topics.


\textbf{Accelerometer} is a three-axis sensor that accurately senses and measures acceleration. 
It is one of the primary sensors embedded into smartphones and has been widely used for gaming, health tracking, and activity recognition\cite{sousa2019human,shen2018closing,hicks2019best}. 
An accelerometer consists of springs, fix electrodes, and an electrode on a movable seismic mass.
When an acceleration is applied along a certain direction, the movable mass moves to the opposite direction, thus changing the capacitance between fixed electrodes. Then the accelerometer can calculate the acceleration by measuring the changed capacitance. In our work, when a built-in speaker plays the audio, it will produce vibrations which will be propagated to the accelerometer via the motherboard.
And the vibrations induce a movement of the accelerometer's mass, registering acceleration. 


Android operating system allows apps to access accelerometer data at various sampling rates. 
By requesting the \textit{SENSOR\_DELAY\_FASTEST} mode\cite{androidApi}, an app can acquire sensor data at the maximum sampling rate. 
However, due to the limitations posed by different smartphone manufacturers, the maximum sampling rate of the accelerometer for this mode can vary between 416$\sim$500Hz\cite{ba2020learning} on modern smartphones (more details in Section~\ref{sec:evalution}). According to $Nyquist\quad sampling \quad theorem$, the accelerometer can only capture the information below $250$Hz while the sampling rate of the accelerometer is $500$Hz. To be able to reconstruct the information in high frequency, we introduce the concept of phonemes.

\textbf{Phonemes} 
are the smallest phonological units divided according to the natural properties of speech\cite{ohala2005phonetic}. 
We take the English language as an example, the phonemes in English are classified into two categories: vowels and consonants\cite{englishphonology}. The number of vowels is $20$ and their energy mainly distributes below $2000$Hz, while the number of consonants is $28$ and their energy mainly distribute below $8000$Hz\cite{baken2000clinical}. However, the accelerometer can only capture limited speech information due to the restricted sampling rate. Fig.~\ref{fig:phoneme} shows the spectrogram of the accelerometer data and corresponding spectrogram of audio for vowels and consonants. 
We can observe there exist unique patterns for each phoneme on the spectrograms of both accelerometer and audio. Based on this observation, we can devise an approach which learns the mappings between the accelerometer data and the audio. Besides, it should have the capability to automatically generate the missing high-frequency components with the low-frequency accelerometer data based on the previously learned mappings.

\begin{figure}[t]
    \centering
    \includegraphics[width=0.9\columnwidth]{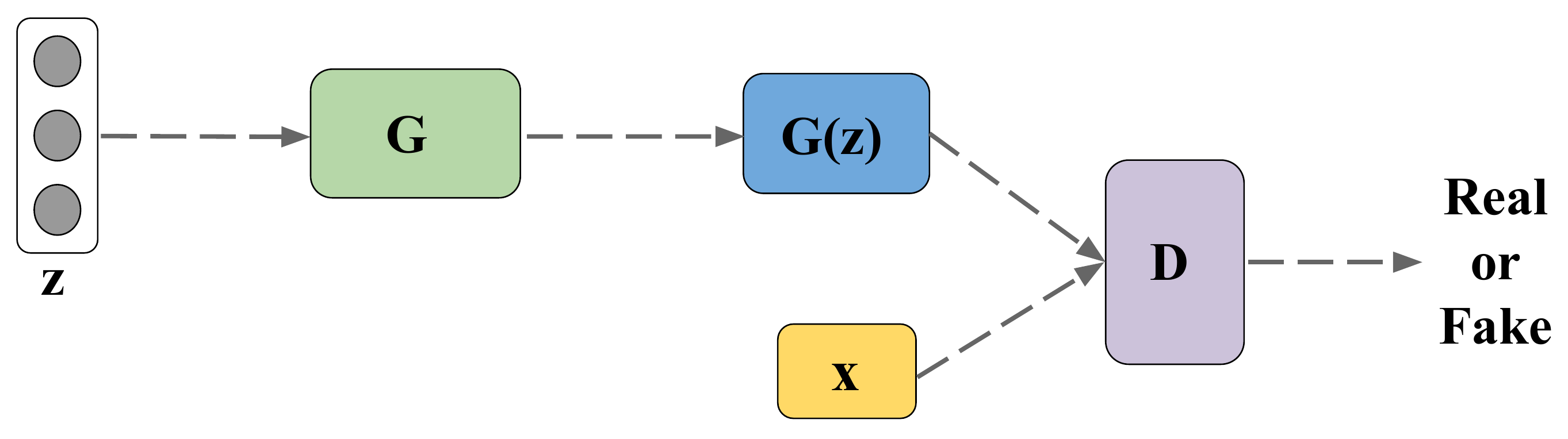}
    \caption{The architecture of Generative Adversarial Networks}
    \label{fig:GAN}
\end{figure}
\begin{figure}[t]
    \centering
    \includegraphics[width=0.9\columnwidth]{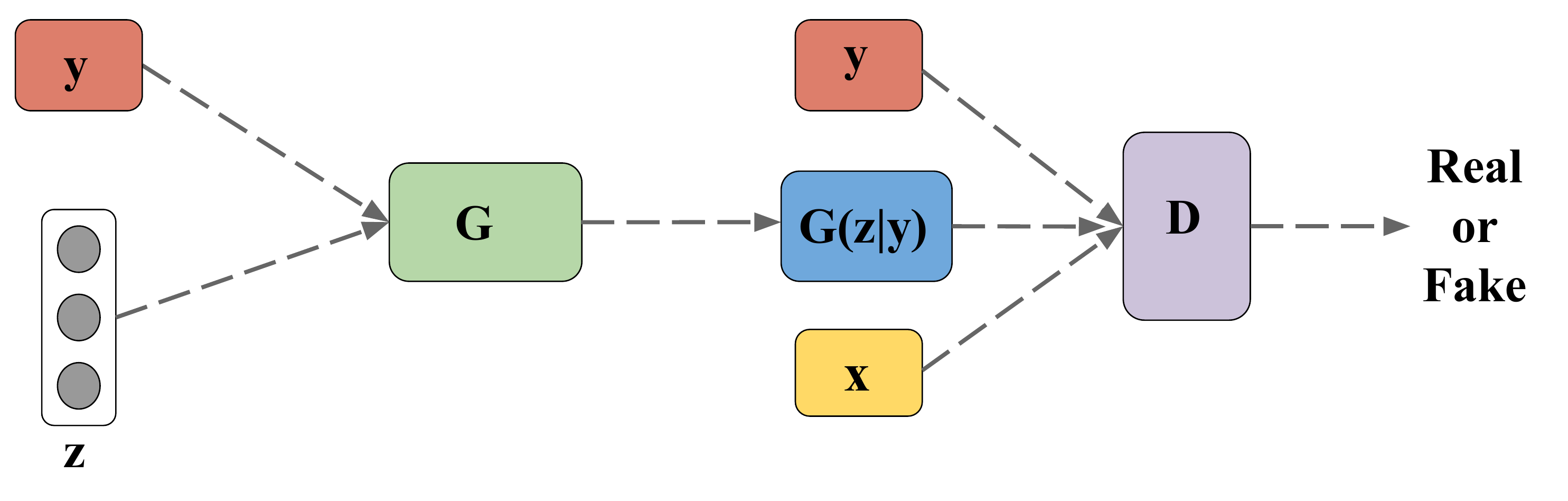}
    \caption{The architecture of conditional Generative Adversarial Networks}
    \label{fig:cGAN}
\end{figure}

\textbf{Generative Adversarial Networks} (GAN)~\cite{goodfellow2014generative} is a machine learning method that engages a game between two neural networks, namely, a generator $G$ and a discriminator $D$. As shown in Fig.~\ref{fig:GAN}, $G$ aims to generate new data
(such as image, music, etc) from a noise vector $z$, while $D$ aims to discriminate the $G(z)$ based on the ground truth $x$. 
During the training process, $G$ constantly evolves to generate new data to try to deceive $D$ as if it is real. Similarly, $D$ also evolves to discriminate the data generated by $G$ as fake. 
The training process terminates until $D$ cannot differentiate between real and the ``fake'' data generated by $G$. 
This implies the data generated by $G$ is indistinguishable from the ground truth. 
However, conventional GANs lack the capability to generate new data that meets desired constraints or conditions. 
A \textbf{conditional GAN} (cGAN)~\cite{2014Conditional}, which architecture is shown in Fig.~\ref{fig:cGAN}, allows us to define a condition $y$ on the input data for a GAN.
Different from the traditional GAN, the generator $G$ aims to generate data $G(z|y)$ from a noise vector $z$ but under the input condition $y$. Besides, the discriminator $D$ still aims to discriminate the generated data from the ground truth $x$. However, $D$ also maps $G(z|y)$ to the original data $x$ via the condition $y$. 
In the training process, G aims to learn such a mapping and generate data that can deceit $D$. 
Therefore, cGAN is a good candidate which can generate the lost high-frequency components based on low-frequency accelerometer data (condition). 


\section{Our audio eavesdropping attack}
\label{sec:system}
\begin{figure*}[h]
    \centering
    \includegraphics[width=0.9\textwidth]{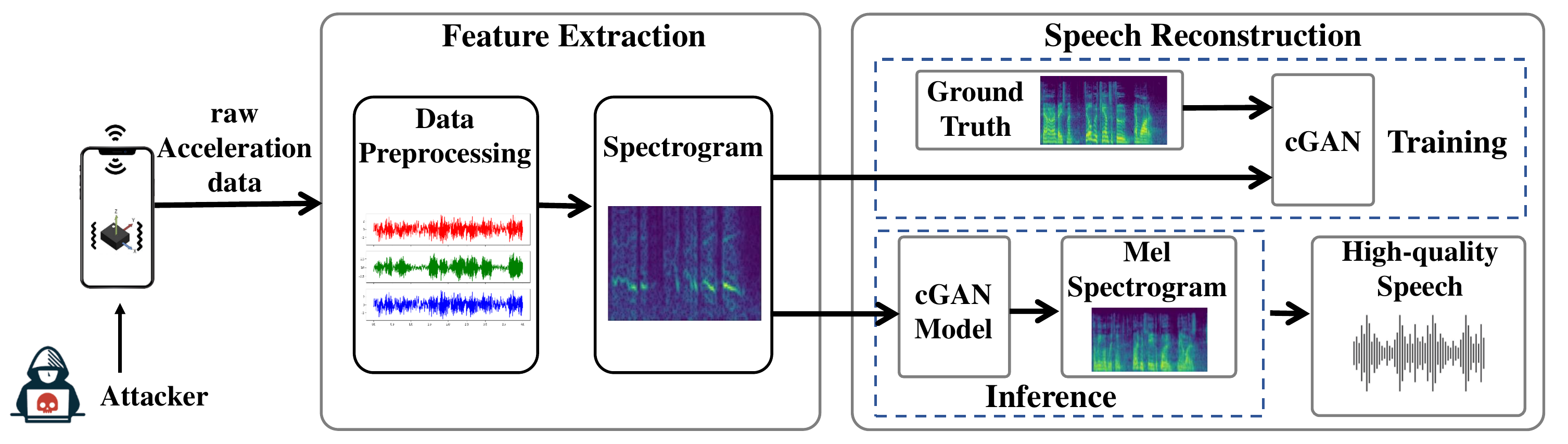}
    \caption{The architecture of \pname system}
    \label{fig:system}
\end{figure*}
In our proposed attack, the accelerometer is used to eavesdrop on the audio played by the built-in speaker on a victim's smartphone. The whole process for the attack and its modules are shown in Fig.~\ref{fig:system}.
In this section, we first define the threat model and assumptions for our attack. 
We then describe in detail the two major components of our attack: feature extraction and speech reconstruction.

\subsection{Threat model}
In our threat model, we assume a spyware has been installed on the victim's smartphone that collects the accelerometer data in the background. 
When the built-in speaker of the victim's phone plays the sound,
the spyware records accelerometer data on all three-axis at the maximum sampling rate in the background. 
Hence, the attacker can access the raw accelerometer data to carry out the eavesdropping attack.   
We only focus on accelerometer data since such sensor has higher sensitivity than the gyroscope, as pointed out by previous research~\cite{ba2020learning}.
Different from the other related works, we assume the attacker has no prior information about the audio playing from the victim speaker, which implies there is no pre-established vocabulary.
It is worth noting that we carry out our attack on the victim's phone independently from internal and external factors. For this reason, we assess its effectiveness under several settings, such as   
the smartphone's manufacturer and model, 
audio output volume from the speaker, 
position (lying on a table or hand-held), 
user movements (still or walking), and 
real-world scenario (e.g. quiet room, restaurant, street).

%


\subsection{Feature Extraction}
In this module, we apply several processing steps to the raw accelerometer data to derive a proper representation as the input for our speech reconstruction module. 

\subsubsection*{\textbf{Zero-mean normalization}} 
The raw accelerometer measurements along $x, y, z$ axis have different baseline value. 
For example, the baseline value of $z$-axis is about $9.8$ due to the earth gravity while the other axes are $0$. 
To exclude the influence of earth gravity, we apply zero-mean normalization to the raw data as follows,
\begin{equation}
\begin{aligned}
    s_{ij}=\frac{s_{ij}-\Bar{s_i}}{\sigma}
\end{aligned}
\end{equation}
where the $s_{ij}$ represents the $j$-th sample of the $i$-th axis, and  $i=1,2,3$ denotes the $x, y, z$ axis respectively, the $\Bar{s_i}$ denotes the mean value of $s_i$, and the $\sigma$ denotes the standard deviation of $s_i$. After zero-mean normalization, the mean value of the data for each axis is zero under stationary scenario.


\subsubsection*{\textbf{High-pass filter}} 
\begin{figure}[t]
\centering
\subfigure[Before High-pass filter]
{\includegraphics[width=0.49\columnwidth]{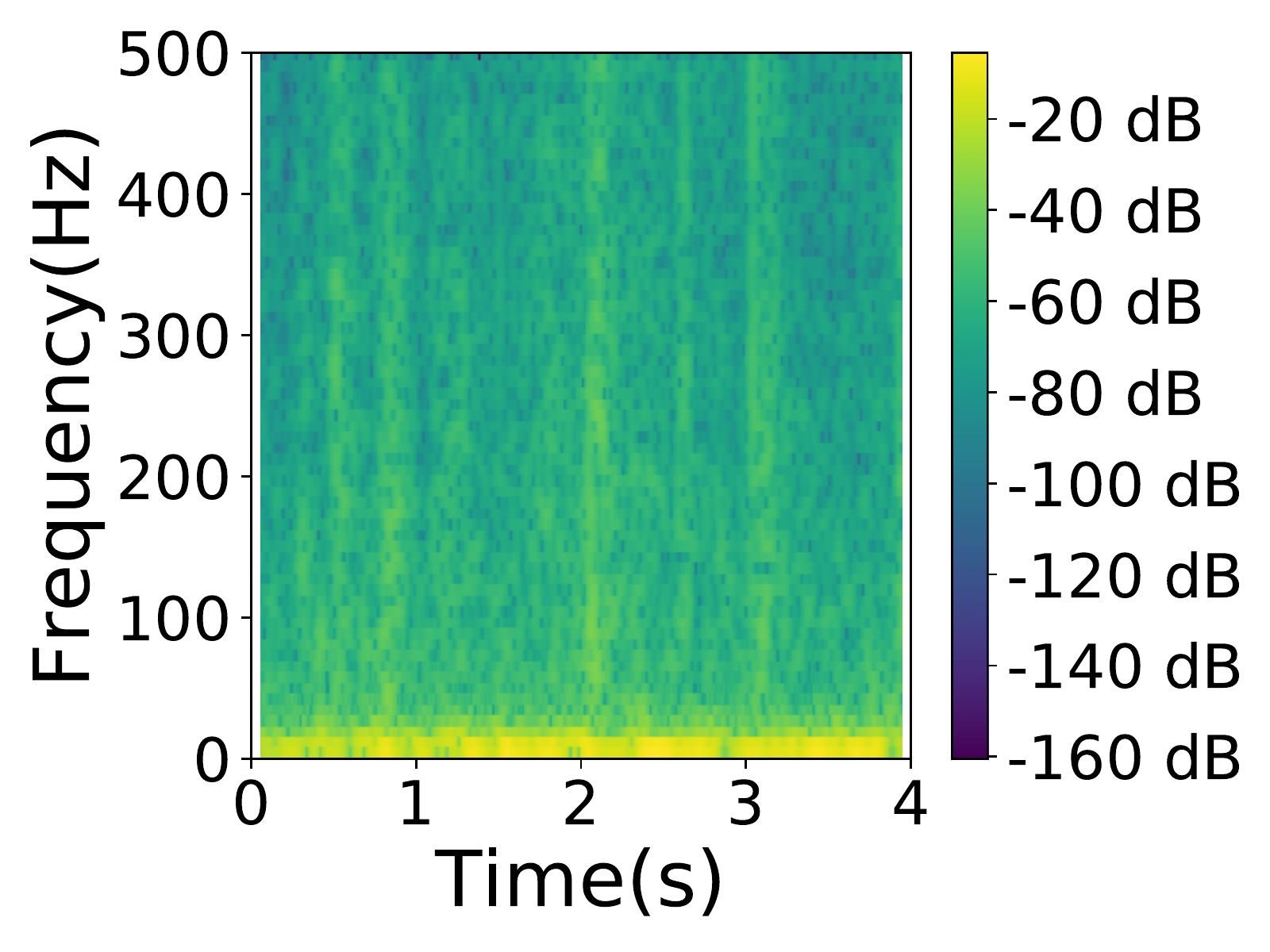}
\label{fig:before}}
\subfigure[After High-pass filter]
{\hspace{-10pt}\includegraphics[width=0.49\columnwidth]{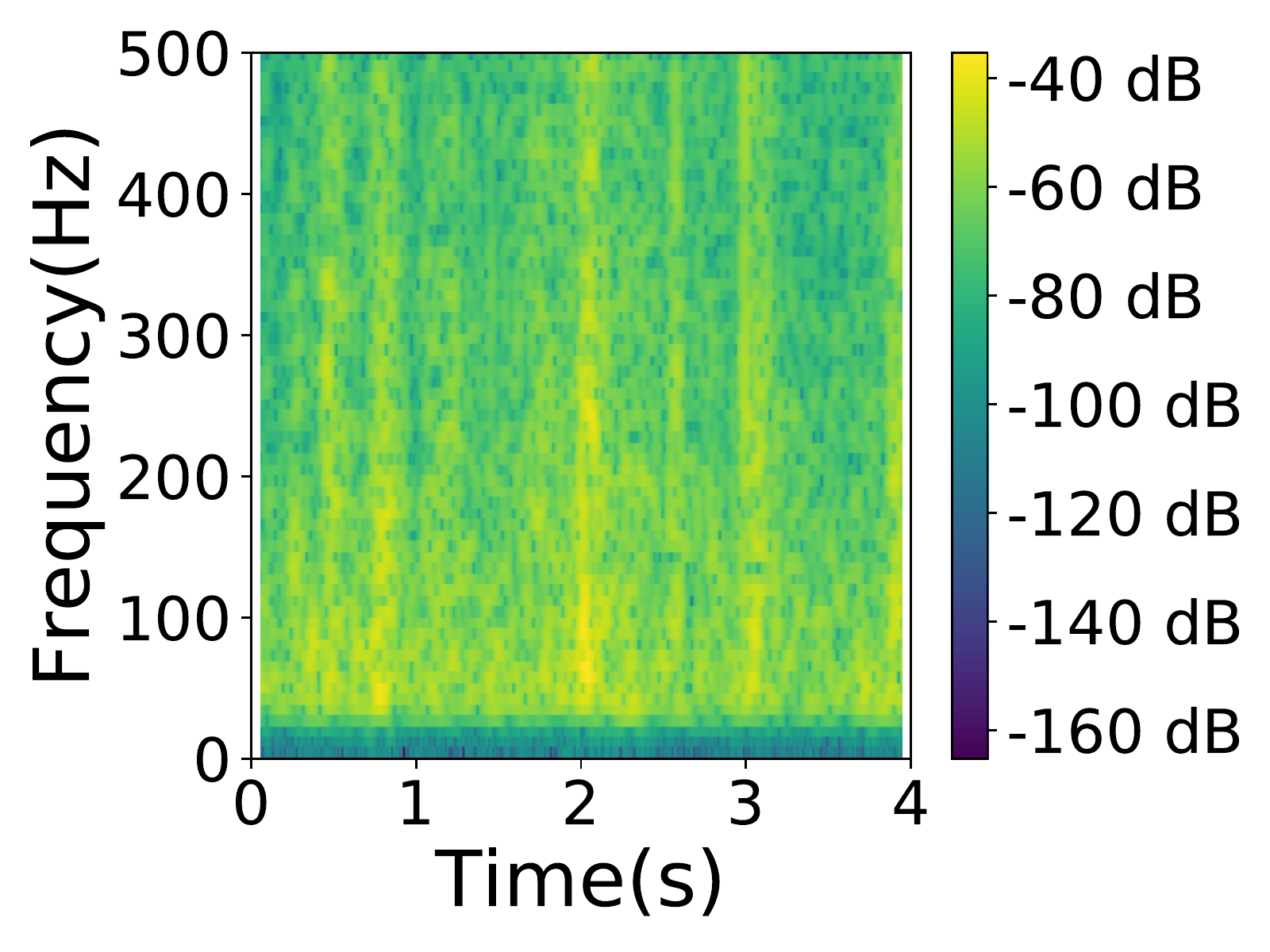}
\label{fig:after}}
\caption{Spectrogram of accelerometer data with human movement}
\hspace{-5pt}
\label{fig:movement}
\vspace{-20pt}
\end{figure}

In real-world scenarios, human activities could significantly influence the accelerometer data.
Fig.~\ref{fig:before} shows the spectrogram of accelerometer data with human movement.
We can observe that the human movement corresponds to a dominant component in the low frequency. 
Hence we use a high-pass filter with a threshold of $20$Hz to remove the impact of human movement\footnote{The fundamental frequency of human speech is above $85$Hz and the perceptible frequency by the human ear is above $20$Hz, the human activities rarely affect the frequency components above $80$Hz~\cite{ba2020learning}.} while preserving as much speech information as possible. 
The spectrogram of the accelerometer signal after applying the high-pass filter is shown in Fig.~\ref{fig:after}. 
The major difference between the original and filtered signals is that the high-frequency speech-related components can be presented clearly after filtering out the low-frequency movement-based components. 

\subsubsection*{\textbf{Interpolation}} 
As mentioned in Section~\ref{sec:preliminary}, the Android operating system provides various sampling rate modes. 
However, the system does not guarantee a fixed time interval between two measurements.
To solve this problem, we apply the linear interpolation approach to the accelerometer data to fill the missing data. 
After interpolation, we obtain a constant sampling rate at $1$kHz for the accelerometer data. 
It is worth noting that while the interpolation fixes the unstable time intervals in the original accelerometer data, it does not introduce extra speech information\cite{ba2020learning}. 

\begin{figure}[t]
\centering
\includegraphics[width=0.9\columnwidth]{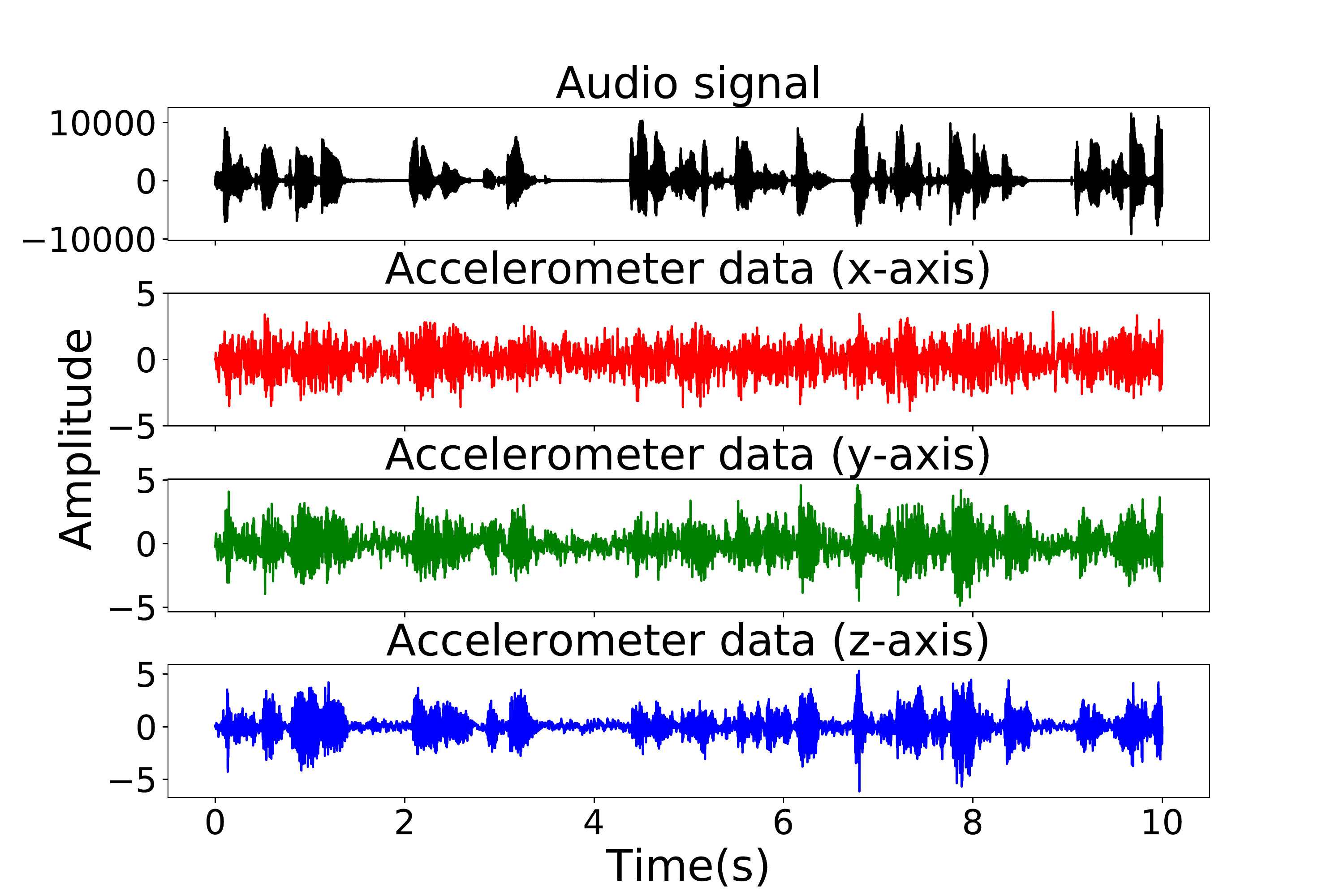}
\caption{
Accelerometer data response to the played audio.
}
\label{fig:signal}
\end{figure} 


\begin{figure*}[ht]
    \centering
    \includegraphics[width=0.8\textwidth]{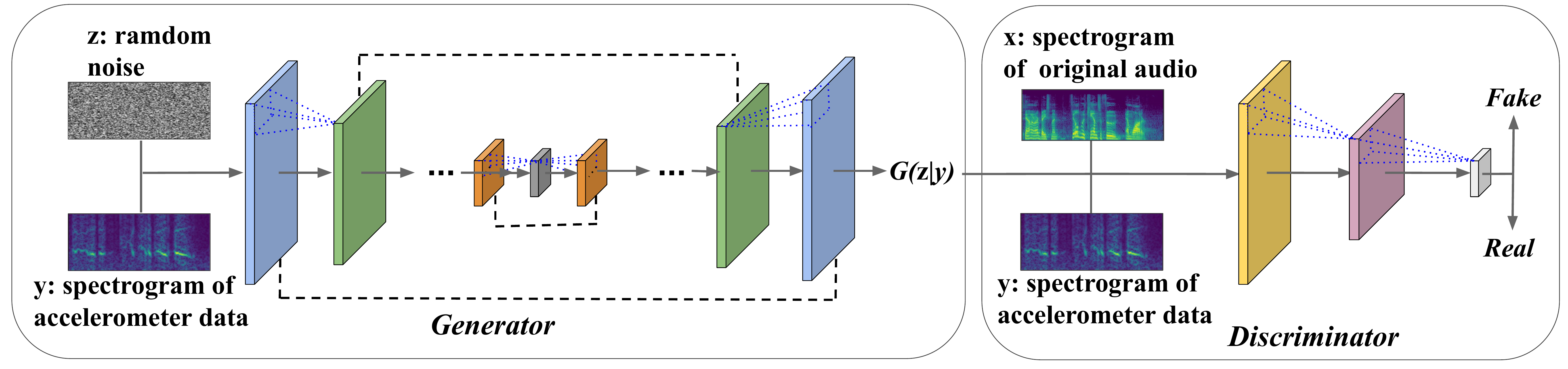}
    \caption{Networks architecture of our conditional Generative Adversarial Network for \pname.}
    \label{fig:network}
\end{figure*}

\subsubsection*{\textbf{Signal-to-spectrogram of Accelerometer data}} 
After the above steps, our accelerometer data is still three temporal signals (one for each axis). 
As the input of cGAN requires a two-dimensional image, we convert the accelerometer data on the most responsive axis to an image-like spectrogram. 

By comparing the waveform of the original audio with the correspondent accelerometer data (as shown in Fig.~\ref{fig:signal}), we can observe that of $z$-axis is more responsive and less noisy than $x$ and $y$ axes. 
Therefore, we choose the $z$-axis accelerometer signal for the next conversion steps. 

We divide the accelerometer signal into the fixed length segments of four seconds 
and apply the Short-Time Fourier transform (STFT) on each segment as follows,
\begin{equation}
\begin{aligned}
&STFT\{s(t)\}(\tau, \omega) \equiv S(\tau, \omega)\\
&\qquad \qquad \qquad \quad \;\;\;  =\int_{-\infty}^{\infty} s(t) w(t-\tau) e^{-i \omega t} d t
\end{aligned}
\end{equation}
where ${\displaystyle w(\tau )}$ is the window function ($Hann$ window is applied in this work), and ${\displaystyle s(t)}$ is the accelerometer data to be transformed. ${\displaystyle S(\tau ,\omega )}$ is the Fourier transform of ${\displaystyle s(t)w(t-\tau )}$ which represents the phase and amplitude of the signal over time and frequency. 

After the STFT, we obtain the spectral characteristics of accelerometer data.
Due to the magnitude of the spectral characteristics is close to zero, we take a square root of the STFT results. 
Then we perform the normalization on the spectral characteristics to speed up the convergence of cGAN in the audio reconstruction module. 


\subsubsection*{\textbf{Audio-to-spectrogram conversion}} 
The audio reconstruction module requires the original audio as ground truth for model training.
Therefore, we also convert the original audio into an image-like spectrogram following a similar process. 
However, different from the above signal-to-spectrogram conversion, we convert the audio signal to a Mel spectrogram. The mathematical relationship between the ordinary frequency scale and the Mel frequency scale can be expressed as follows\cite{stevens1937scale},
\begin{equation}
\operatorname{Mel}(f)=2595 * \log _{10}(1+f / 700)
\end{equation}
where $f$ refers to the frequency.
This conversion is necessary since the perception in a human ear is not linear in terms of frequency.
In particular, the human ear is more sensitive to low frequencies than high frequencies~\cite{kumar2019melgan}. 
The Mel scale~\cite{stevens1937scale} is the nonlinear transformation of frequency which distorts the original audio frequency for better human perception. 

\subsection{Speech Reconstruction}
The purpose of eavesdropping is to reconstruct the original audio via the accelerometer data. We adopt a GAN variant 
to enhance the spectrogram of the accelerometer data via the generation of the high-frequency features, which are absent from such signal.

\subsubsection*{\textbf{conditional Generative Adversarial Networks (cGAN)}} 
As we mentioned above, traditional GAN can only generate the new data close to the training samples from random noise. However, our main purpose is to transform the spectrogram of accelerometer data to the Mel spectrogram of corresponding audio. To enable the model to generate the corresponding Mel spectrogram according to the different spectrogram of accelerometer data, we refer the conditional GAN approach and take the spectrogram of accelerometer data as the condition.

Fig.~\ref{fig:network} illustrates our network architecture of cGAN. The input for our cGAN is the ground truth $x$ (i.e., the Mel spectrogram of original audio) and the condition $y$ (i.e., the spectrogram of accelerometer data). From the combination of a noise vector $z$ and condition $y$, the generator $G$ generates $G(z|y)$ as one of the inputs for the discriminator $D$. 
Additionally, the ground truth $x$ and the condition $y$ are combined as another input of $D$, which represents the real image under condition $y$. 
During the joint training process, $D$ tries to discriminate the $G(z|y)$ from the ground truth $x|y$ while $G$ tries to adjust its network parameters to generate a $G(z|y)$ which can fool $D$. 
For each phoneme in a word, $G$ automatically learns the mapping from accelerometer data spectral features to speech spectral features through the zero-sum game between $G$ and $D$. Once the training process completed, the generator $G$ can correctly reconstruct a word pronunciation via the accelerometer data, even if the word does not appear in our training set.


\subsubsection*{\textbf{Objective}} To enable our reconstructed audio more closely to the original audio, we define the loss function of magnitude spectrogram of generated audio signals and original audio signals\cite{ananthabhotla2019towards}. It can be expressed as
\begin{equation}
L_{S}=\left\|S(t, f)-S_{p}(t, f)\right\|_{1}, t \in T, f \in F
\end{equation}
where  $S(t, f)$ and $S_{p}(t, f)$ are the magnitude spectrogram representation of the generated audio signals and original audio signals respectively.

According to cGAN\cite{2014Conditional}, the generator $G$ aims to minimize $\log (1-D(G(\boldsymbol{z} \mid \boldsymbol{y})))$ while discriminator $D$ aims to maximize $\log (1-D(G(\boldsymbol{z} \mid \boldsymbol{y})))$, as if they are following the two-player min-max game. The objective of the cGAN is as follows.
\begin{equation}
\begin{aligned}
    \min _{G} \max _{D} V_{cGAN}(D, G)=&\mathbb{E}_{\boldsymbol{x}}[\log D(\boldsymbol{x} \mid \boldsymbol{y})]+\\
   & \mathbb{E}_{\boldsymbol{z}}[\log (1-D(G(\boldsymbol{z} \mid \boldsymbol{y})))]
\end{aligned}
\end{equation}
where x is the ground truth, y is the condition, and z is the noise prior.
Combining the loss function of signals magnitude and the objective of conditional GAN, our final objective is 
\begin{equation}
\begin{aligned}
     &L^{*}=\left\|S(t, f)-S_{p}(t, f)\right\|_{1}+\\
     &\qquad \;\;  \mathbb{E}_{\boldsymbol{x} \sim p_{\text {data }}(\boldsymbol{x})}[\log D(\boldsymbol{x} \mid \boldsymbol{y})]+\\
   &\qquad \;\; \mathbb{E}_{\boldsymbol{z} \sim p_{z}(\boldsymbol{z})}[\log (1-D(G(\boldsymbol{z} \mid \boldsymbol{y})))], t \in T, f \in F
\end{aligned}
\end{equation}
\subsubsection*{\textbf{Generator Architecture}} Traditional Encoder-Decoder network in generator needs all information flows to pass through all layers. However, in the image to image translation problems, inputs and outputs are shared on the low-level information that does not need to be considered for conversion\cite{isola2017image}. Therefore, it will increase the calculated costs and time costs if we adopt the traditional Encoder-Decoder network. To address this problem, we use U-Net\cite{ronneberger2015u} as the network architecture of the generator. The whole U-Net architecture is symmetrical, layers on the left are convolutional layers and on the right are upsampling layers. The convolutional layers extract the feature with square kernels of size $4\times4$ and stride value 2, and when the image passes a convolutional layer, its size will be changed. The upsampling layers predict the pixel label by decoding the feature. Different from the traditional Encoder-Decoder network, the feature maps obtained from each convolutional layer are concatenated to the corresponding upsampling layer so that the feature maps of each layer can be effectively used in subsequent calculations, this is known as skip connections (the gray dashed line in the left panel of Fig.\ref{fig:network}). 

\subsubsection*{\textbf{Discriminator Architecture}} Our discriminator has three convolutional layers. Different from the general discriminator, we only discriminate the image at the scale of patches instead of the entire image. It tries to classify each $30\times 30$ patch in an image as real or fake. 
At the end of the training process, the output of $D$ is the average of all the responses from a convolutional pass across the image. 

\subsubsection*{\textbf{Training}}  
We train each individual user model with 200 epochs. In the first 100 epochs, we set the learning rate of 0.0002, and in the last 100 epochs, we use Adam~\cite{adam} to adaptive adjust the learning rate to speed up the convergence of the network. The detailed algorithm for the training process is presented in Algorithm~\ref{alg:training}, where $\theta_{D}$ and $\theta_{G}$ represent the parameters (such as weights, bias, etc.) of generator $G$ and discriminator $D$ respectively, $m$ refers to the batch size, $x^{i}$ refers to the ground truth, $y^{i}$ refers to the condition, $z^{i}$ refers to the noise sample. In each iteration, we first fixed the parameters of generator $\theta_G$ and update the parameters of discriminator $\theta_D$, after $\theta_{D}$ updated, we will keep $\theta_{D}$ fixed and update $\theta_{G}$.

\begin{algorithm}[h]
\small
\caption{Training Process of cGAN}
\label{alg:training}
\hspace*{0.02in} {\bf Input:}
$n$ paired training data \\  $\left\{\left(y^{1}, x^{1}\right),\left(y^{2}, x^{2}\right), \ldots,\left(y^{n}, x^{n}\right)\right\}$\\
\hspace*{0.02in} {\bf Output:} $\theta_{D}$, $\theta_{G}$
\begin{algorithmic}[1]
\FOR{each epoch}
    \FOR{each iteration}
    \STATE Sample $m$ paired examples from input  

    \STATE Sample $m$ noise samples $\left\{z^{1}, z^{2}, \ldots, z^{m}\right\}$ from a distribution.
    \STATE Generate data $\left\{\tilde{x}^{1}, \tilde{x}^{2}, \ldots, \tilde{x}^{m}\right\},  \tilde{x}^{i}=G\left(y^{i}|z^{i}\right)$ 
    \STATE Update discriminator parameter $\theta_{D}$ to maximize 
\begin{equation*}
    \begin{aligned}
         &\tilde{V}=L_S+\frac{1}{m} \sum_{i=1}^{m} \log D\left( x^{i}|y^{i}\right)+\\
&\qquad\frac{1}{m} \sum_{i=1}^{m} \log \left(1-D\left( \tilde{x}^{i}|y^{i}\right)\right),\\
& \theta_{D} \leftarrow \theta_{D}+\eta \nabla \tilde{V}\left(\theta_{D}\right)
    \end{aligned}
\end{equation*}
    \STATE Sample $m$ noise samples $\left\{z^{1}, z^{2}, \ldots, z^{m}\right\}$ from a distribution.
    \STATE Sample $m$ conditions $\left\{y^{1}, y^{2}, \ldots, y^{m}\right\}$ from input
    \STATE  Update generator parameter $\theta_{G}$ to maximize 
    \begin{equation*}
        \begin{aligned}
             &\tilde{V}=\frac{1}{m} \sum_{i=1}^{m} \log \left(D\left(G\left( z^{i}|y^{i}\right)\right)\right),\\
             &\theta_{G} \leftarrow \theta_{G}-\eta \nabla \tilde{V}\left(\theta_{G}\right)
        \end{aligned}
    \end{equation*}
    \ENDFOR
\ENDFOR
\end{algorithmic}
\end{algorithm}

\begin{table*}[ht]
\centering 
\begin{tabular}{llcccccc}
\hline
\textbf{Label} & \textbf{Person} & \textbf{Sex} & \textbf{Language} & \textbf{Length(seconds)} & \textbf{Testing words} & \textbf{Training words} & \textbf{Overlapping words} \\ \hline \hline
User$_1$ & Bill Gates & male & English & 7068 &179&12593&19\\ \hline
User$_2$ & Feifei Li & female & English & 7120 &182&17626&15\\ \hline
User$_3$ & Pony Ma & male & Chinese & 5180 &215&28554&20\\ \hline
User$_4$ & Jane Goodall & female & English & 7484 &188& 11339 & 23 \\ \hline
User$_5$ & Jiaying Ye & female & Chinese & 9032 & 188 & 11339 & 16 \\ \hline
User$_6$ & Mingzhu Dong & female & Chinese & 5428 &234&18709&22\\ \hline
User$_7$ & Steve Job & male & English & 14836 & 190&37751&17 \\ \hline
User$_8$ & Yansong Bai & male & Chinese & 6792 & 251& 27317& 22 \\ \hline
User$_9$ & Anne Hathaway & female & English & 60 & 197 & * & 21\\ \hline
User$_{10}$ & Elon Musk & male & English & 60 &156& * &17 \\ \hline
User$_{11}$ & Mark Zuckerberg & male & English & 60 & 177 & * & 15 \\ \hline
User$_{12}$ & Oprah Winfrey & female & English & 60 & 167 & * & 18 \\ \hline
User$_{13}$ & Lan Yang & female & Chinese & 60 & 289 & * & 25 \\ \hline
User$_{14}$ & Minhong Yu & male & Chinese & 60 & 199 & * & 17 \\ \hline
User$_{15}$ & Robin Li & male & Chinese & 60 & 244 & * & 20 \\ \hline
User$_{16}$ & Yingtai Long & female & Chinese & 60 & 198 & * & 18\\ \hline
\end{tabular}
\caption{The dataset used for evaluating \pname, and note that, each audio couples with the accelerometer signal. Data of the last $8$ users are used to evaluate the performance of cross-users.}
\label{tab:users}
\end{table*}

\subsubsection*{\textbf{Spectrogram-to-audio conversion}} After obtaining the Mel spectrogram generated by conditional GAN, we need a vocoder to convert the acoustic parameters to speech waveform. In our system, we adopt a classic vocoder Griffin-Lim\cite{griffin1984signal} to synthesize the waveform from the Mel spectrogram.  
The Griffim-Lim algorithm is a method to reconstruct the speech waveform with a known amplitude spectrum and an unknown phase spectrum by iteratively generating the phase spectrum and using the known amplitude spectrum and the calculated phase spectrum. 
We first initialize a phase spectrum and synthesize a new speech waveform with this phase spectrum and a known amplitude spectrum (from the Mel spectrogram generated by cGAN) by Short-time Fourier Inverse Transform (ISTFT). 
Then, we perform STFT to the new speech waveform and calculate the new phase spectrum. 
We continue to synthesize the new speech waveform with the known amplitude spectrum and the new phase spectrum until we  obtain the satisfactory waveform.
\section{Evaluation}
\label{sec:evalution}

\begin{table}[h]
\setlength{\tabcolsep}{6mm}
\centering
\begin{tabular}{cl}
\hline
   \textbf{Score}  & \hspace{1.5cm} \textbf{Level} \\ \hline \hline
   5 & Recovered all of the original speech \\ \hline
   4 & Recovered most of the original speech \\ \hline
   3 & Recovered half of the original speech \\ \hline
   2 & Recovered little of the original speech \\ \hline
   1 & Recovered none of the original speech \\ \hline
\end{tabular}
\vspace{5pt}
    \caption{MOS and corresponding level.}
    \label{tab:mos_score}
\end{table}


In this section, we report the details of our experimental setup and performance evaluation of \pname on the reconstruction of speech from accelerometer data.

\subsection{Implementation and Experiment Setup} 
In our experiments, we target smartphones running the Android operating system since its prevalent share on the smartphone market, i.e., $72.21\%$ reported by Statista\cite{statista}. 
In this work, we evaluate our attack scheme with multiple sampling rates to accommodate both the legacy and future permission policies of the Android system\cite{ratelimit}.
We collect accelerometer data from six different smartphones (Huawei Mate40 Pro, Huawei Mate30 Pro, OPPO Reno6 Pro, Samsung S21+, OPPO Find X3, and XiaoMi RedMi 10X Pro) and two different tablets (Huawei MatePad Pro and Samsung Galaxy Tab S6 Lite) using a third-party application named \textit{Accelerometer Meter}\footnote{Accelerometer Meter v1.32 - \url{https://keuwl.com/Accelerometer/}} by Keuwlsoft. 
We provide the detailed parameters of these devices in Table~\ref{tab:para} in Appendix~\ref{appendix:parametersmodels}. The highest sampling rate of such smartphones is around $500$Hz. 

We perform both the pre-process of the accelerometer data and conversion of the enhanced accelerometer Mel spectrogram back into audio on a laptop with an i7-10750H CPU and 16GB memory. The training and testing processes run on a server with Nvidia RTX 3090 GPU. 
We train an individual model for each public personality by using his/her audio samples respectively and then train several generic models with the data of a specific group of personalities. For each model, we train it in $200$ epochs with the initial learning rate of $0.002$. A model training process takes about $2.28$ hours on a dataset with $1010$ Mel spectrogram images.



\begin{figure*}[ht]
    \centering
    \includegraphics[scale=0.36]{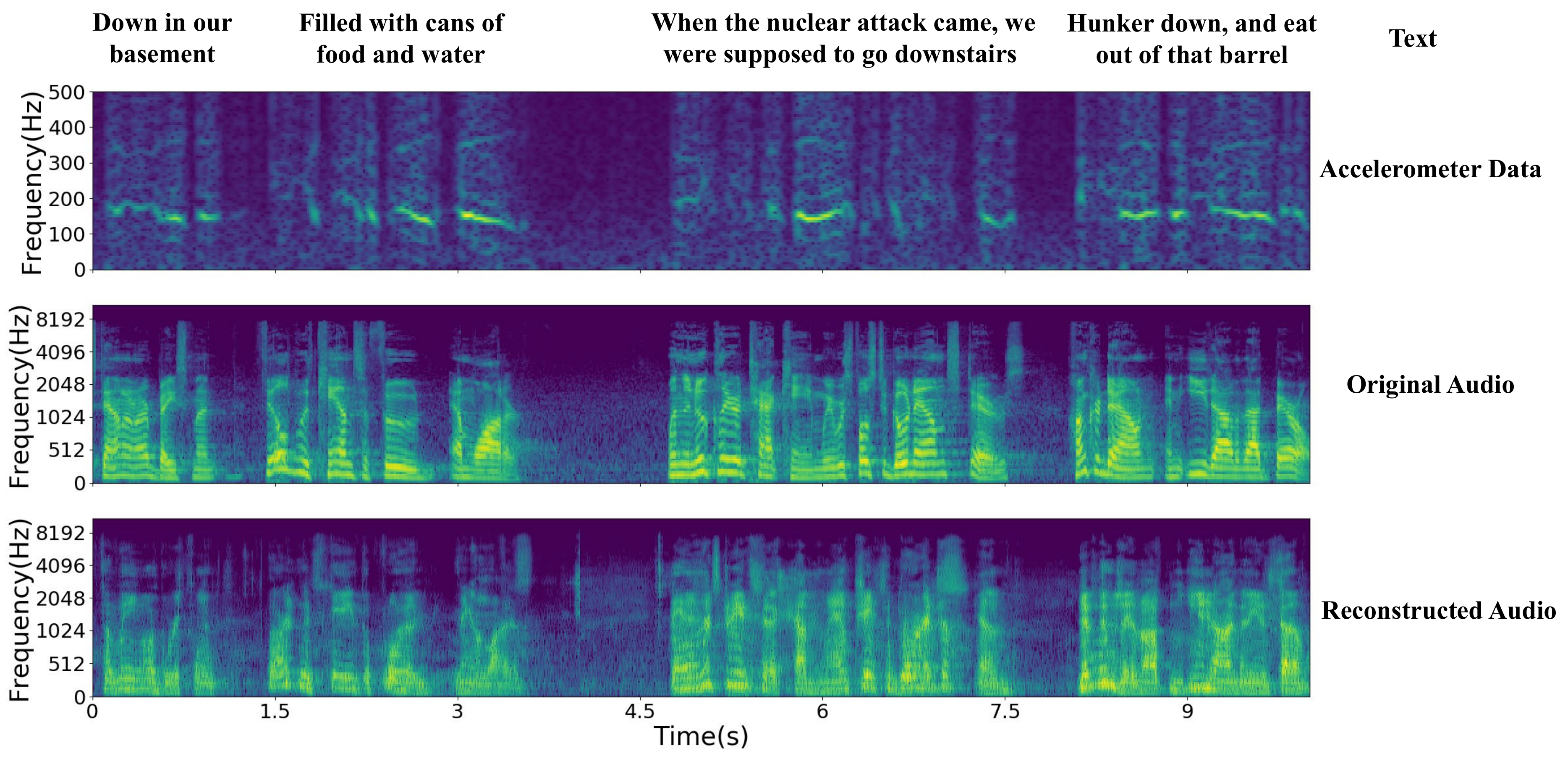}
    \caption{User$_1$ speech spectrograms for (a) accelerometer data, (b) original audio and (c) reconstructed audio via \pname.}
    \label{fig:Mel}
\end{figure*}

\subsection{Data Collection}

\subsubsection*{\textbf{Audio Collection}}
We collected the audio samples from 8 English-speaking and 8 Chinese-speaking public personalities whose utterances are available on the Internet (e.g., YouTube). 
For convenience, we marked the above public personalities as User$_{1}$ to User$_{16}$ as shown in Table~\ref{tab:users}.
The speech samples of each user are divided into training and testing sets which include different numbers of words\footnote{https://github.com/hui-zhuang/AccEar.git}. 
To demonstrate the effectiveness of reconstructing unlimited words, we make sure that the training and testing sets overlap only on a small set of words.

\subsubsection*{\textbf{Accelerometer Data Collection}}
We put the smartphones on the table in a conference room and play the above collected audio samples with a built-in loudspeaker while our app runs in the background to record the accelerometer data. 
Thus, we have a direct correspondence between audio samples and accelerometer data.
The accelerometer data is divided into training and testing sets coupled with audio samples as shown in Table~\ref{tab:users}.
In addition, we verify the robustness of \pname by collecting accelerometer data under different settings (i.g., sampling rates, volume, phone models, position, scenarios).





\subsection{Evaluation Metrics}
To evaluate the performance of reconstructed audio, we adopt the following three metrics.

\textbf{Mel-Cepstral Distortion (MCD)}~\cite{tamamori2017speaker}
is an objective evaluation metric since it represents the difference of the Mel-Frequency Cepstral Coefficients (MFCC) features between the reconstructed audio and the corresponding original audio.
Therefore, a small MCD means that the reconstructed audio is similar to the original one (i.e., the smaller, the better).
Typically, reconstructed audio with MCD below $8$ can be comprehended by a speech recognition system~\cite{yan2019feasibility}.
The MCD can be calculated as:
\begin{equation}
M C D=\frac{10}{\log 10} \sqrt{2 \sum_{m=1}^{M}\left(c_{r}(m)-c_{s}(m)\right)^{2}}
\end{equation}
where $c_{r}$ and $c_{s}$ are the Mel-Cepstrum from the original and reconstructed audio, respectively, and $M$ is order of Mel-Cepstrum.

\textbf{Mean Opinion Score (MOS)}\cite{mos} is a subjective evaluation metric
for measuring the intelligibility of the reconstructed audio. 
We recruited twenty volunteers to assess the reconstructed audio on the test set. 
These participants include both native English and Chinese speakers (equal number of female and male) with ages from 20 to 30 years old. All of them are at least with bachelor degree, and they
were all informed of the purpose of our experiments.
To avoid any bias, they participate voluntarily in our experiments without any compensation, and we do not have any incentives.
We ask the participants to first listen to the reconstructed audio and then the original audio immediately after. 
They rate the similarity between the reconstructed and original audio on a scale from $1$ to $5$ as reported in Table~\ref{tab:mos_score}. 
For example, the volunteers give a score of $5$ if they think that the reconstructed audio completely sounds like the original audio. Conversely, they give a score of $1$ if they consider that the reconstructed speech is not at all similar to the original speech.

\textbf{Word Error Rate (WER)} is a commonly used metric in speech recognition to evaluate the accuracy of word recognition. In order to keep the recognized word sequence consistent with the ground truth word sequence, some words need to be substituted, deleted, or inserted (i.e., incorrectly recognized words). WER is the percentage of the number of error words divided by the total number of words in the standard word sequence. It can be calculated as follows
\begin{equation}
 WER = \frac{S+D+I}{N}\times 100\%
\end{equation}
where $S$, $D$, $I$, and $N$ represent the number of substitutions, deletions, insertions, and total words in the standard word sequence, respectively. We recruited 20 volunteers to listen to the original and the reconstructed audio, and recognize the words. Then, we calculate the WER through the words sequences from original and reconstructed audios. A lower WER corresponds to a better comprehensibility of the reconstructed audio.

\subsection{\textbf{Overall Performance Evaluation}} 
We play the audios from the test set on a Huawei Mate40 Pro placed on the table and collect the corresponding accelerometer data. 
Subsequently, we preprocess the accelerometer data to generate the spectrogram. 
After preprocessing, we input the generated spectrogram to the models trained by individual user data or a specific group of users' data.   
And then we get the Mel spectrogram of reconstructed speech and convert it to the audio, and finally we calculate the MCD, MOS and WER.

To report the results more intuitively, we first plot the three types of spectrograms for User$_1$: accelerometer data, original audio, and audio reconstructed from accelerometer data via our cGAN model. 
In Fig.~\ref{fig:Mel}, we can observe that the spectrograms of original audio and reconstructed audio show high similarity. This indicates that our cGAN model is able to learn how to enhance the accelerometer spectrograms by adding specific acoustic components at high frequencies. 
Since the words overlap between training and testing 
sets are small (see Table~\ref{tab:users}), \pname can work on unconstrained vocabulary. 


\begin{figure}[t]
    \centering
    \includegraphics[scale=0.28]{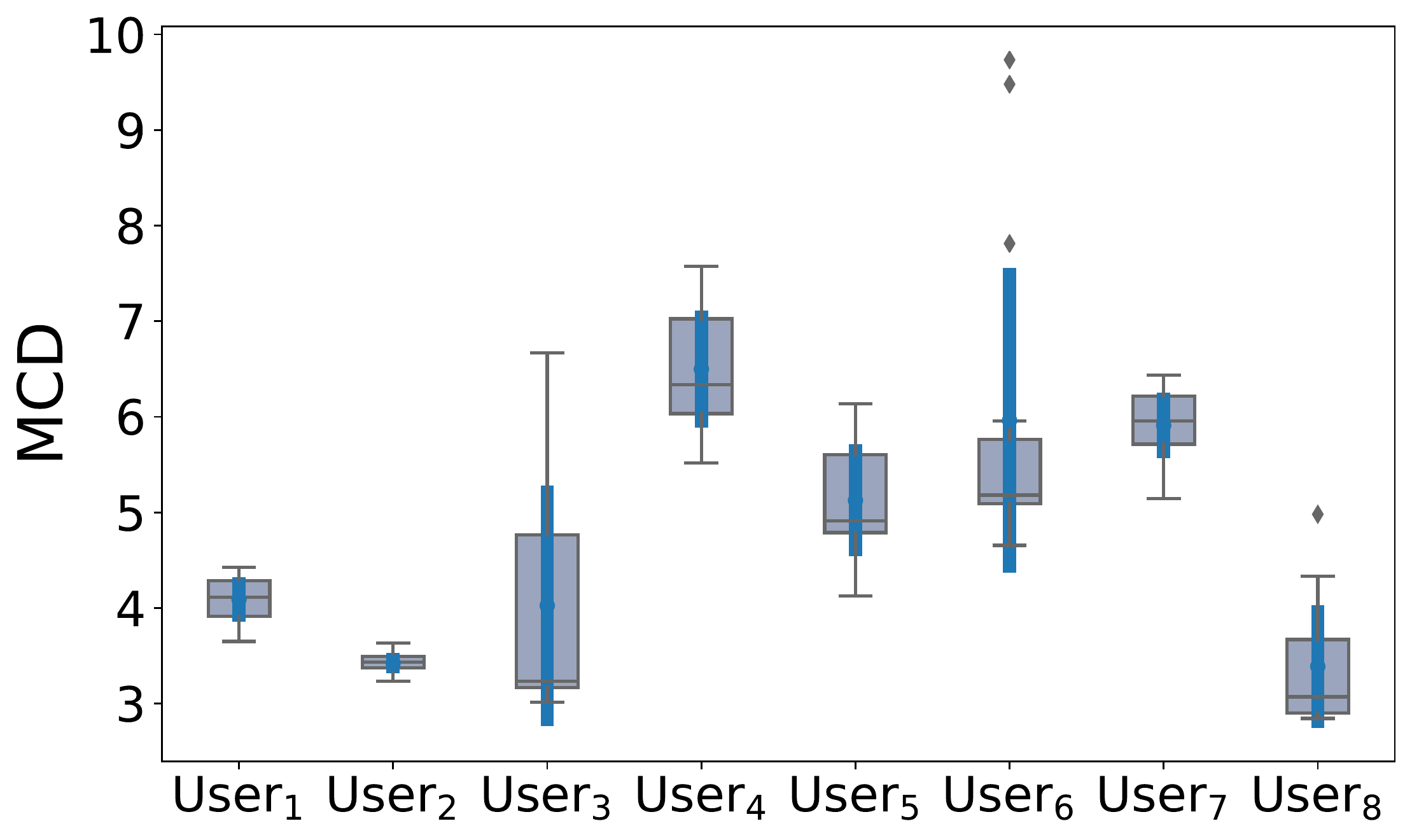}
    \caption{Objective assessment based on MCD for the reconstructed audio}
    \label{fig:base_mcd}
\end{figure}

\begin{figure}[t]
    \centering
    \includegraphics[scale=0.28]{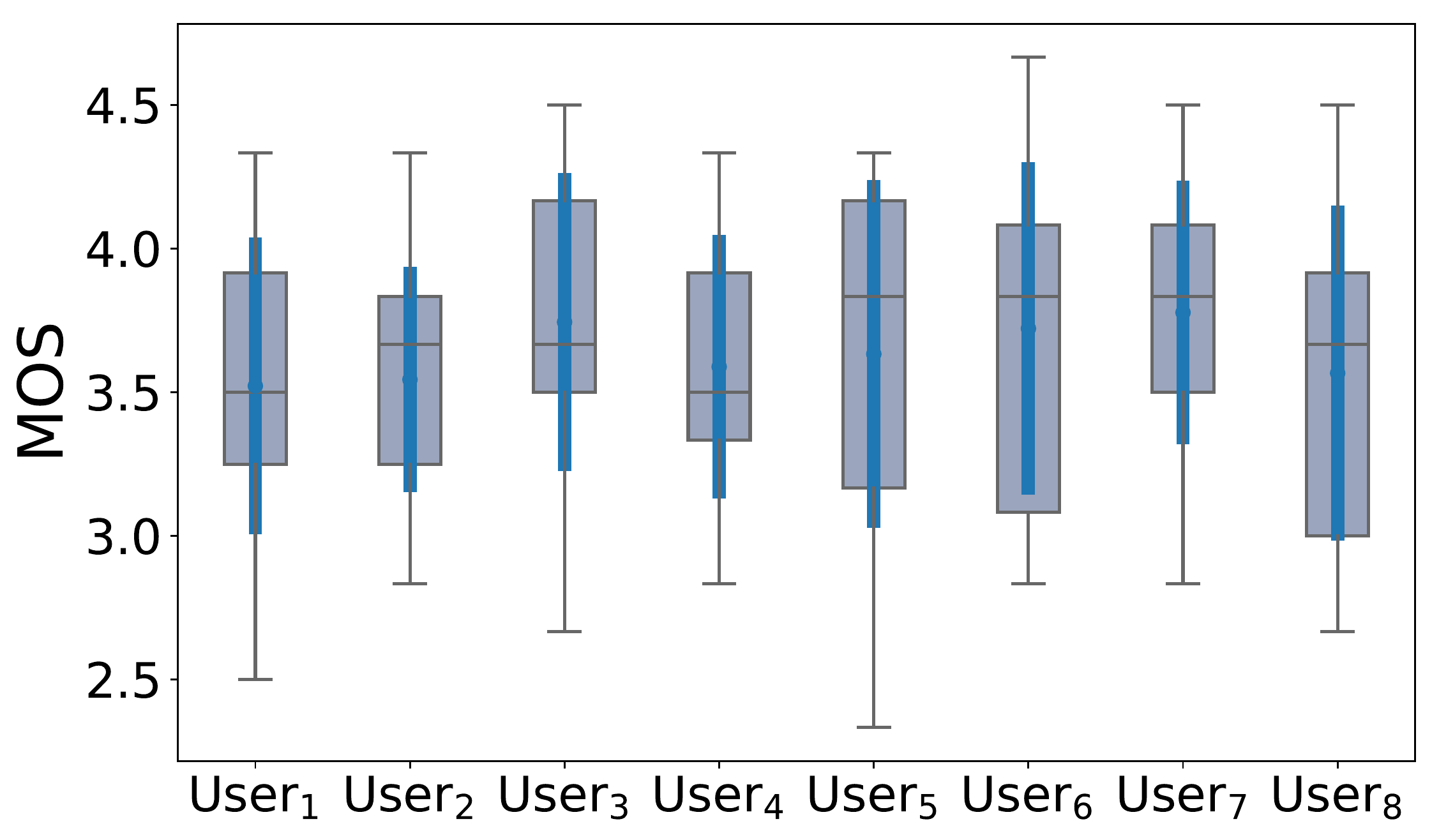}
    \caption{Subjective assessment by volunteers for the reconstructed audio}
    \label{fig:mos}
\end{figure}

\begin{figure}[t]
    \centering
    \includegraphics[scale=0.28]{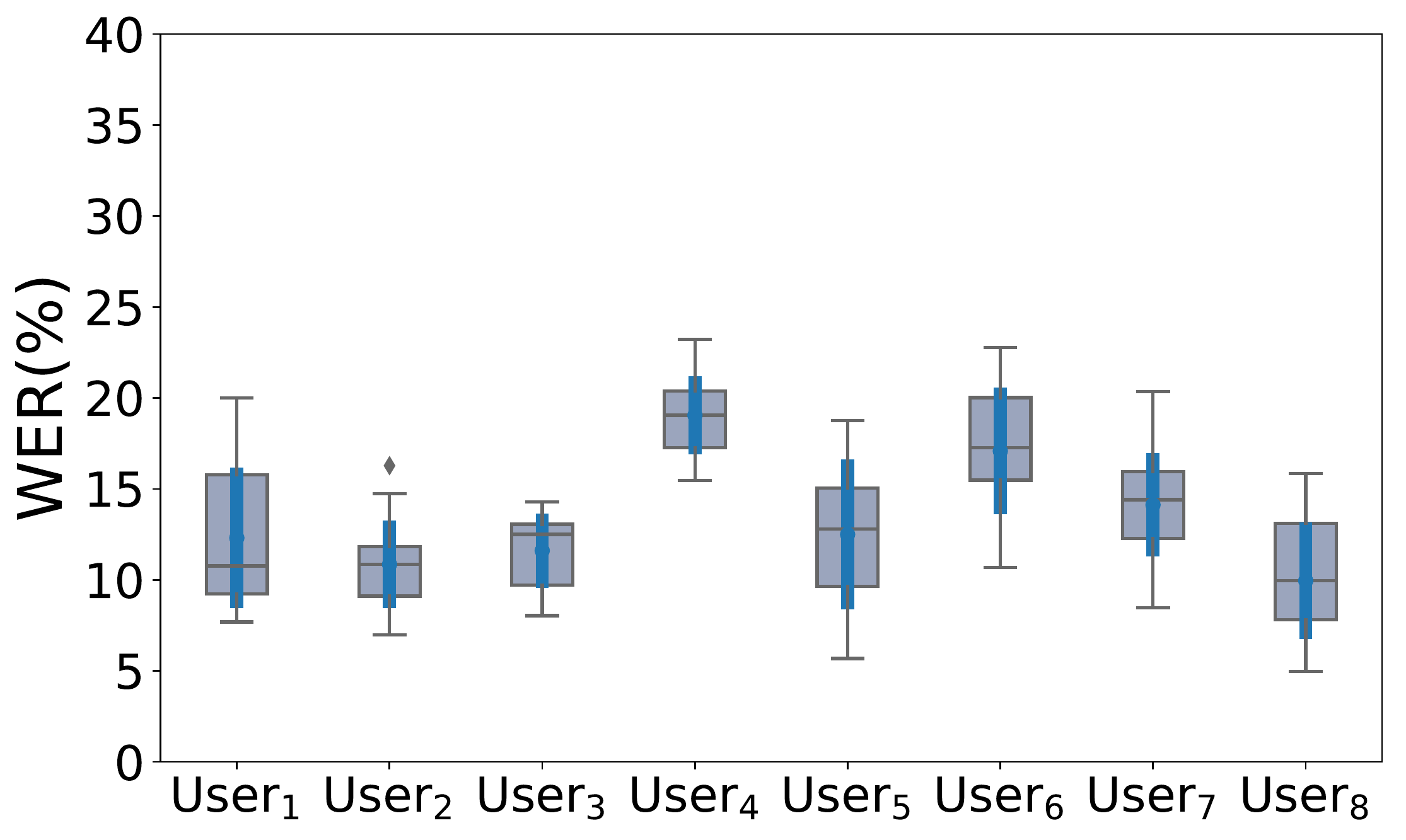}
    \caption{Word Error Rate based on volunteer recognition for the reconstructed audio}
    \label{fig:WER}
\end{figure}

As each individual's pronunciation has unique features, we train an individual model for each one of the top 8 users to better grasp their voice characteristics. 
Fig.~\ref{fig:base_mcd} and Fig.~\ref{fig:mos} illustrate the detailed distribution of MCD and MOS for each individual model. Among the box-plot figures, the $i-$th endpoint on the broken line represents the mean $m_i$ of the data in the $i-$th box, and the blue bold line on each box represents the range from $m_i - std_i$ to $m_i + std_i$, where $std_i$ represents the standard deviation of the $i-$th box.
For the evaluation based on MCD in Fig.~\ref{fig:base_mcd}, we can observe that almost all of the samples have a value lower than $8$, except for several abnormal samples on the model of User$_6$. 
In Fig.~\ref{fig:mos}, we can notice that almost every model has three-quarters of the samples with MOS values above $3$. 
We evaluate the comprehensibility of the reconstructed audio using WER. As shown in Fig.~\ref{fig:WER}, we observe that the average WER of all models are lower than $20\%$, and the average WER of the User$_8$ model is even lower than $10\%$, which indicates that our model can reconstruct the words with high accuracy. These results of MCD, MOS and WER validate that the reconstructed audio is similar to the original audio in terms of waveform, human hearing perception, and word-level comprehensibility, respectively. We also randomly select some reconstructed samples in Table~\ref{tab:mcd} in Appendix~\ref{appendix:mcdrecontruct} to show the relation between MCD and comprehensibility.


\begin{figure*}[h]
\centering
\subfigure[Diversity of speaking the same word]
{\includegraphics[width=0.6\columnwidth]{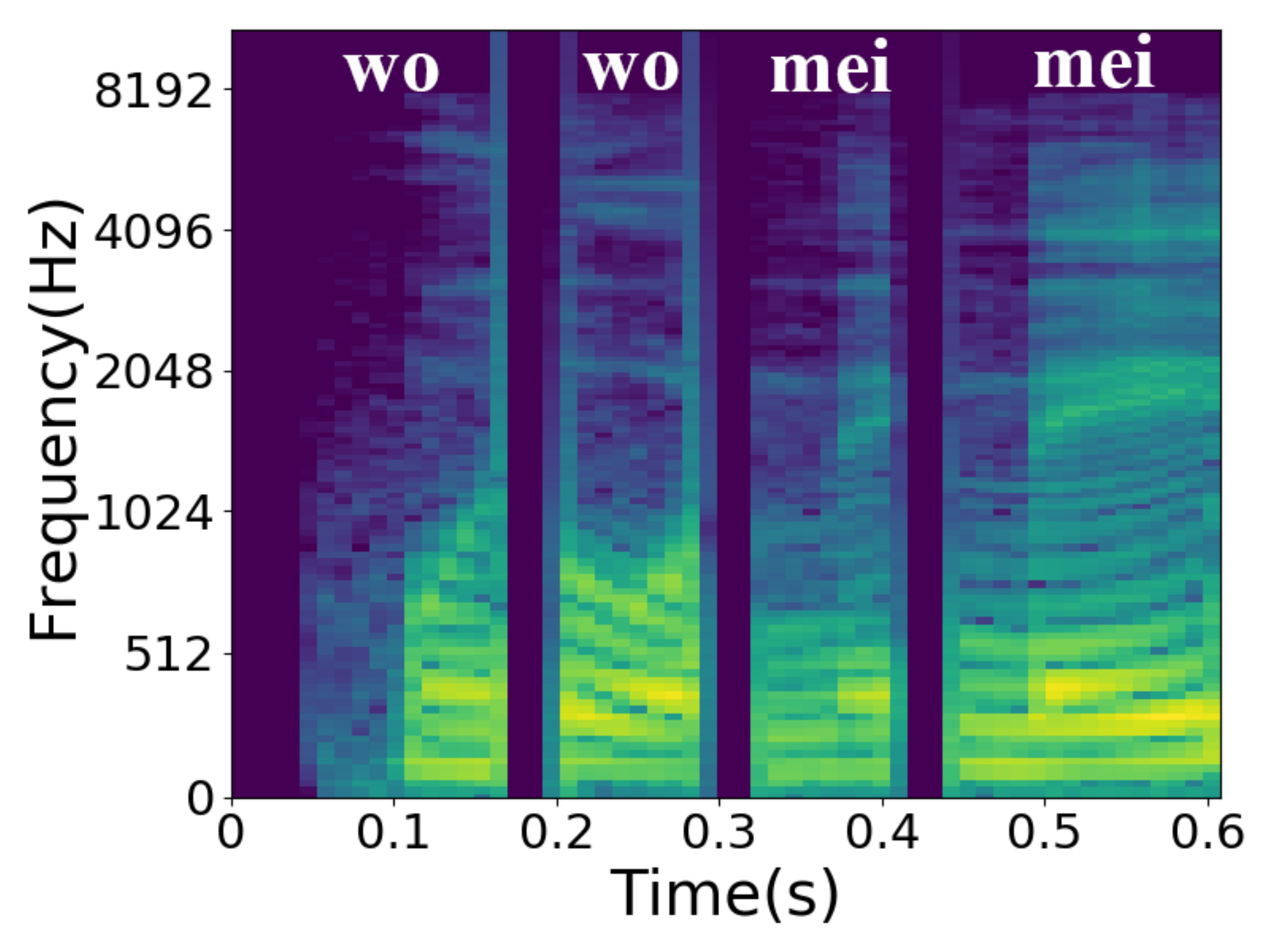}
\label{fig:variation_high}}
\subfigure[Original audio]
{\includegraphics[width=0.6\columnwidth]{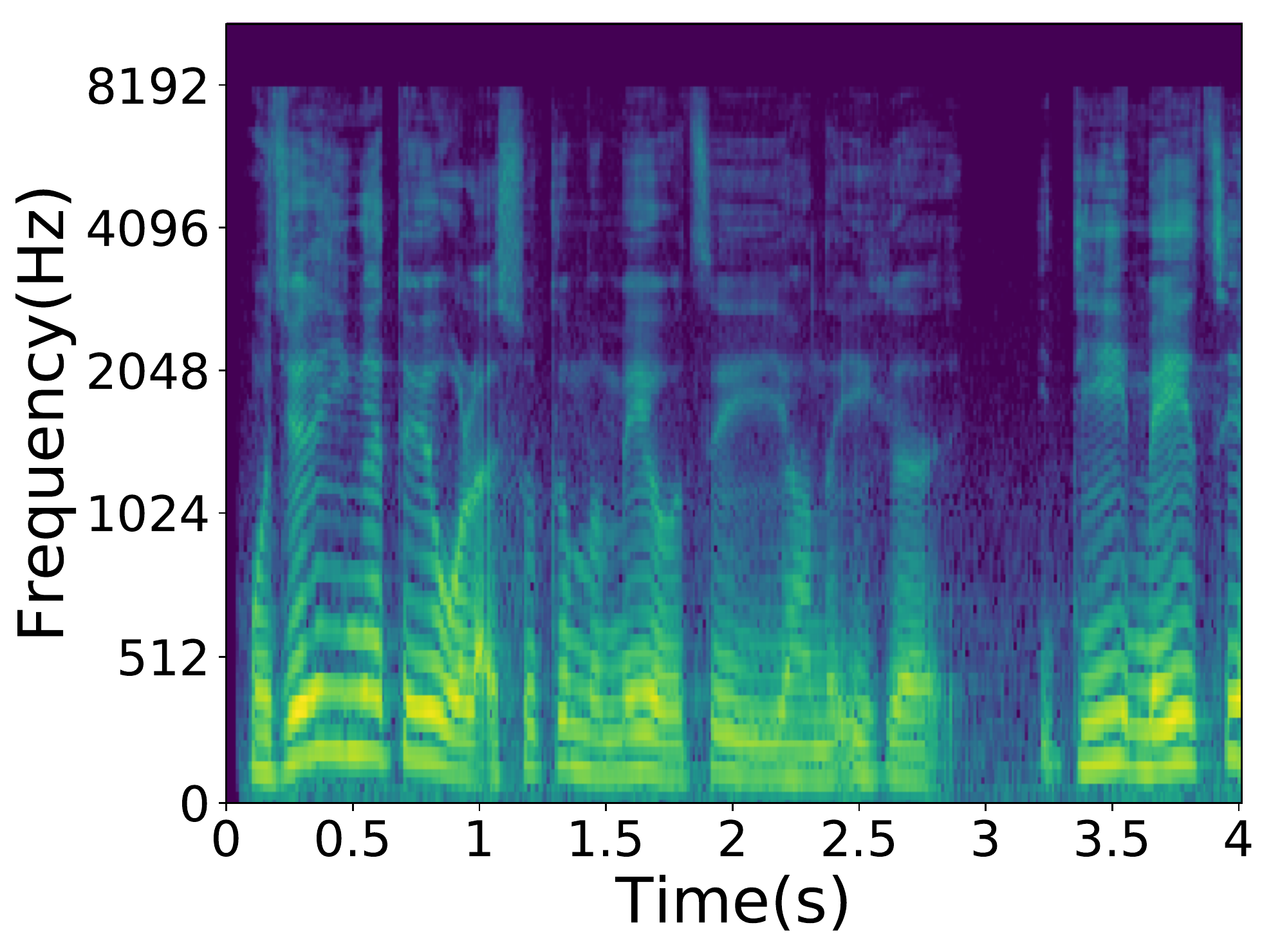}
\label{fig:dmz_original}}
\hspace{-5pt}
\subfigure[Reconstructed audio]
{\includegraphics[width=0.6\columnwidth]{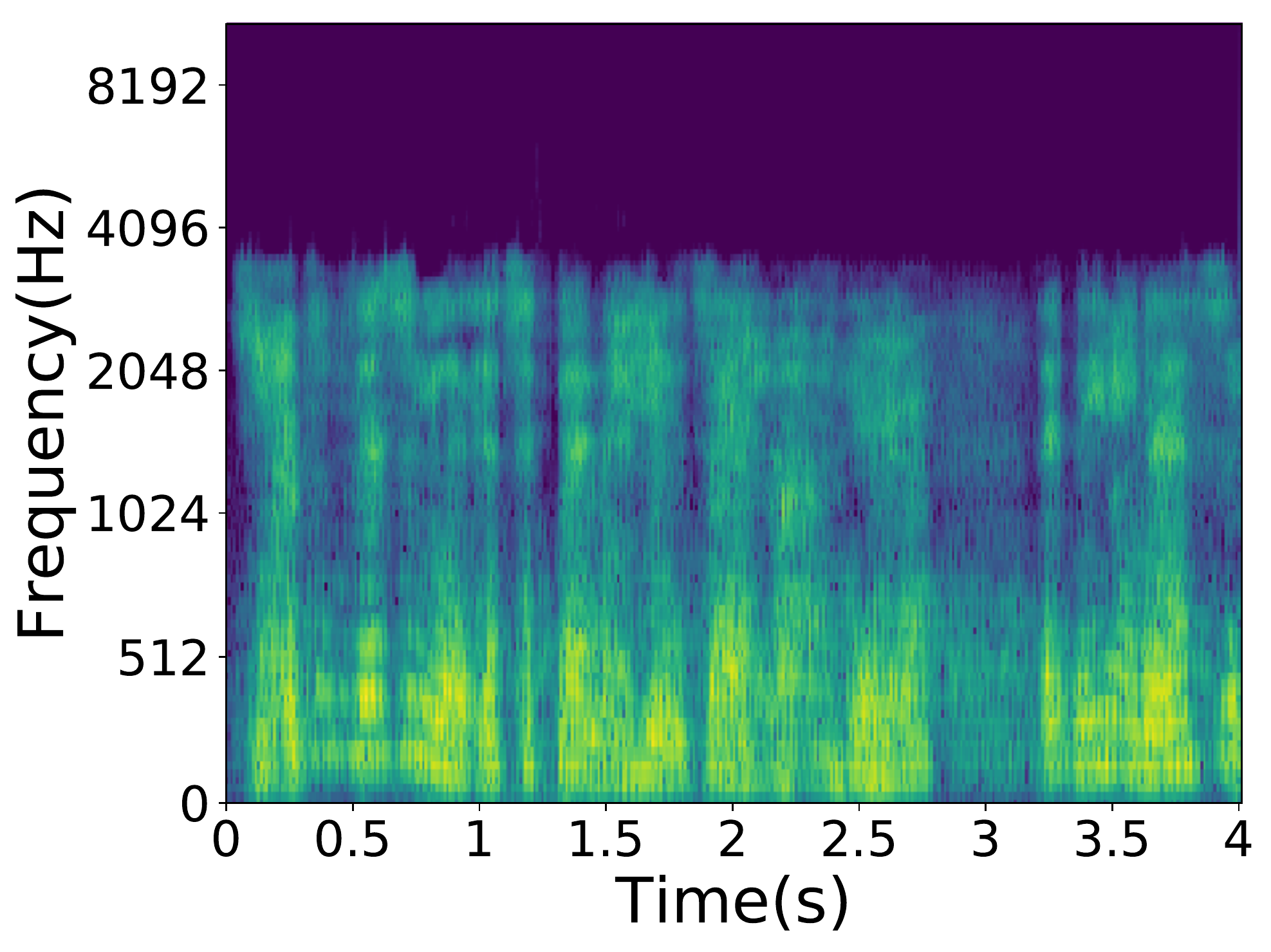}
\label{fig:dmz_reconstructed}}
\caption{The Mel spectrogram of User$_6$. The variation of the same word is large for User$_6$. The original audio and reconstructed audio show high similarity in the low-frequency region but the high-frequency components of reconstructed audio are missing.}
\label{fig:dmz}
\end{figure*} 

\begin{figure*}[h]
\centering
\subfigure[Diversity of speaking the same word.]
{\includegraphics[width=0.6\columnwidth]{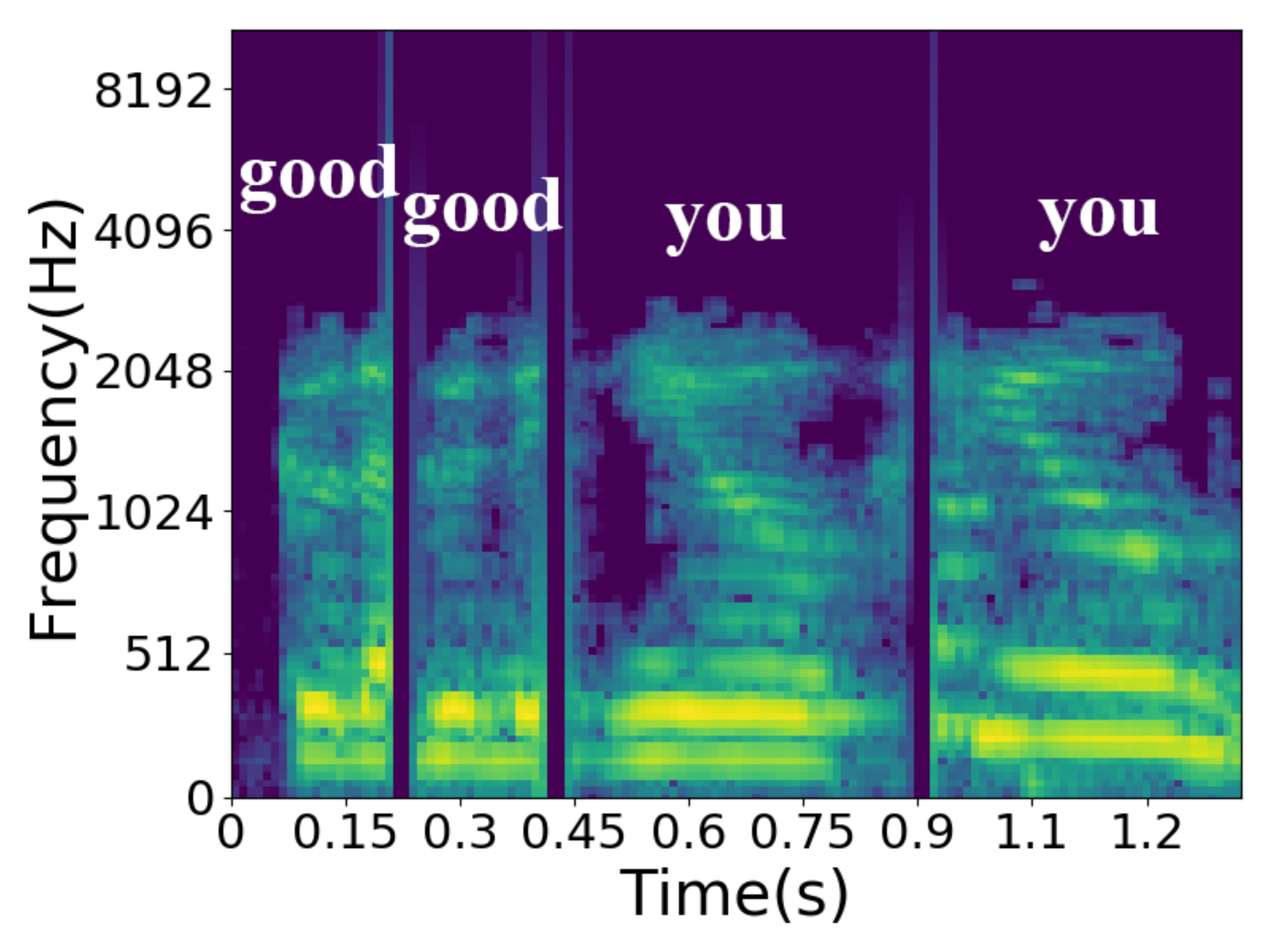}\label{fig:variation_low}}
\subfigure[Original audio.]
{\includegraphics[width=0.6\columnwidth]{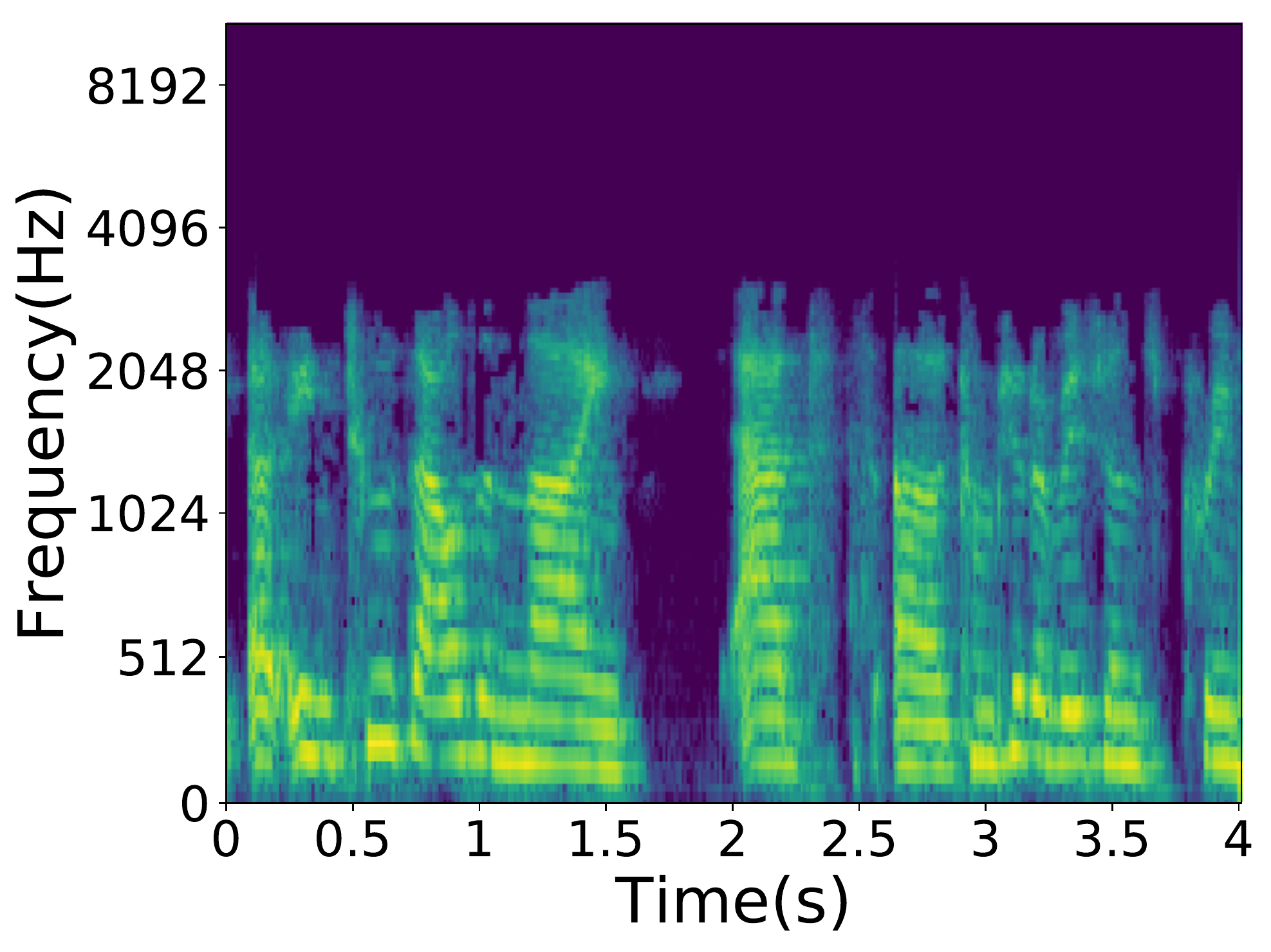}
\label{fig:lff_original}}
\hspace{-5pt}
\subfigure[Reconstructed audio.]
{\includegraphics[width=0.6\columnwidth]{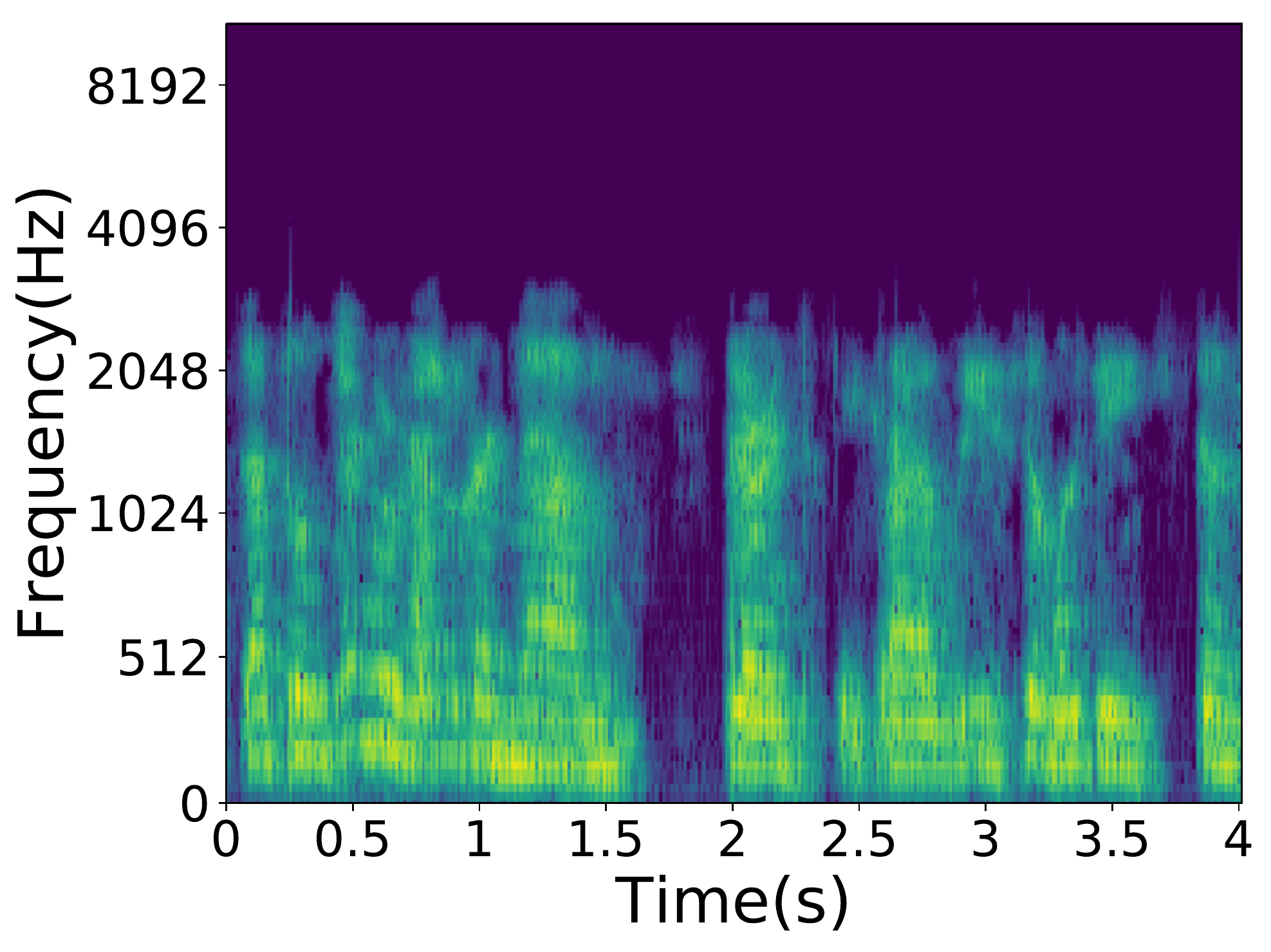}
\label{fig:lff_reconstructed}}
\caption{The Mel spectrogram of User$_2$. The variation of the same word is small for User$_2$. As the high-frequency components of User$_2$ are less than User$_6$, the audio can be reconstructed more accurately.}
\label{fig:lff}
\end{figure*}

We further investigate the outliers observed on the model of User$_6$. Fig.~\ref{fig:variation_high} delineates the pronunciation diversity of the same words, and Fig.~\ref{fig:dmz_original} depicts User$_6$ has wide vocal spectrum with the frequency range of 0$\sim$8000Hz. 
We can notice that the reconstructed audio is similar to the original audio in the low frequency components, but the high frequency components are not reconstructed as expected, which results in a high MCD.

That is because the sampling rate of accelerometer data is only $500$Hz in this case, even though our model can infer the high frequency components according to the signature of low frequency components, it is difficult to fully recover the high-frequency components when the variation of a phoneme is large for a person with wide vocal frequency spectrum as shown in Fig.~\ref{fig:dmz}.


In addition, we observe that the reconstruction for original audios with relatively low frequency have better performance since there is less pronunciation diversity. For example, the major frequency of User$_2$ is below $4096$Hz, as shown in Fig.~\ref{fig:lff_original}. The spectrogram variation of the same word (Fig.~\ref{fig:variation_low}) is much less than the user with wider vocal spectrum, so the Mel spectrogram of original audio and reconstructed audio of User$_2$ are highly similar which is also verified by the corresponding MCD, MOS, and WER scores. We further 
discuss the influence of the diversities in Appendix~\ref{appendix:coverageeffect}.
\begin{figure*}[t]
\centering
\subfigure[Volume]
{\includegraphics[width=0.23\textwidth]{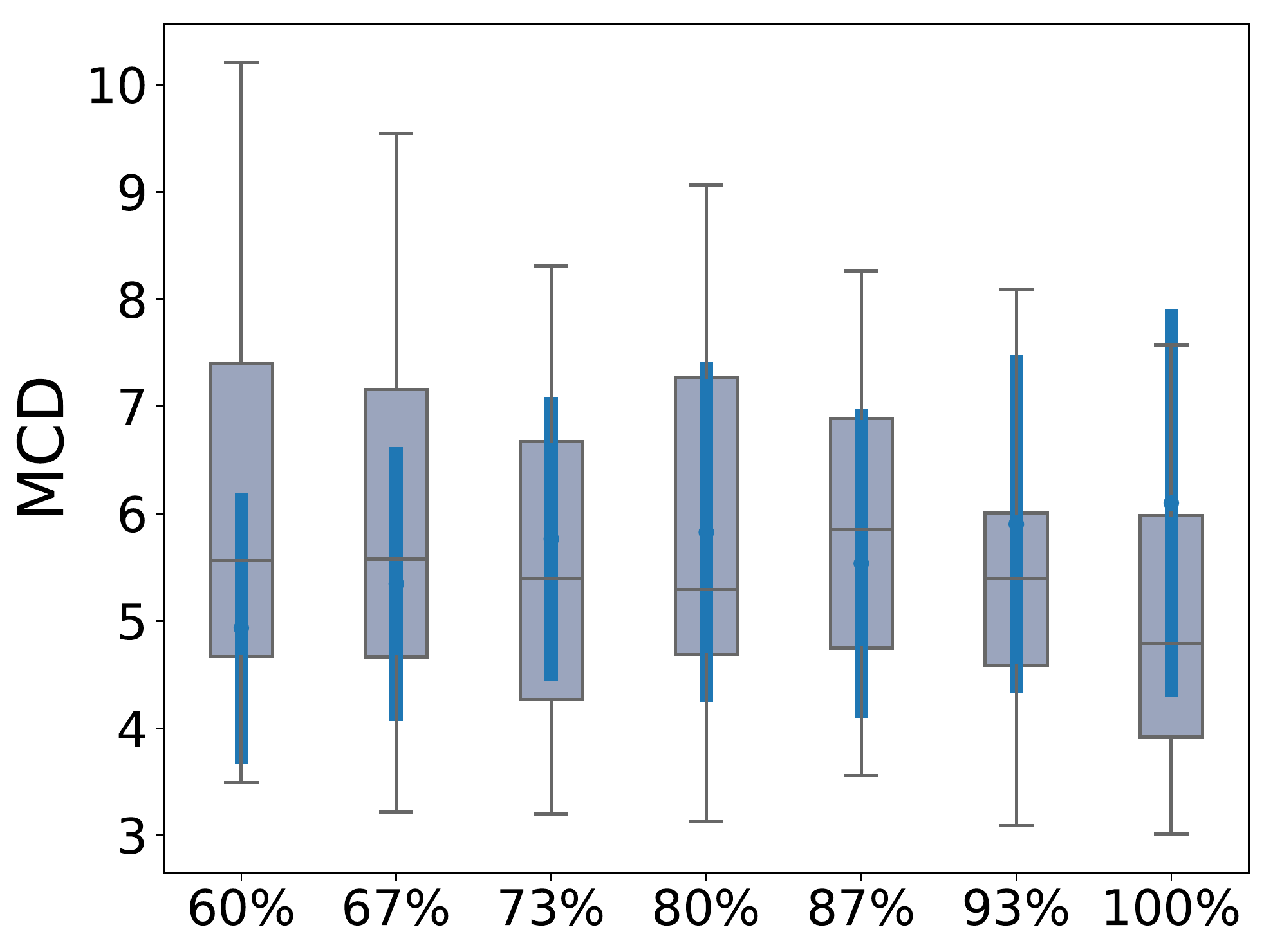}
\label{fig:volume}
}
\subfigure[Placement]
{\includegraphics[width=0.23\textwidth]{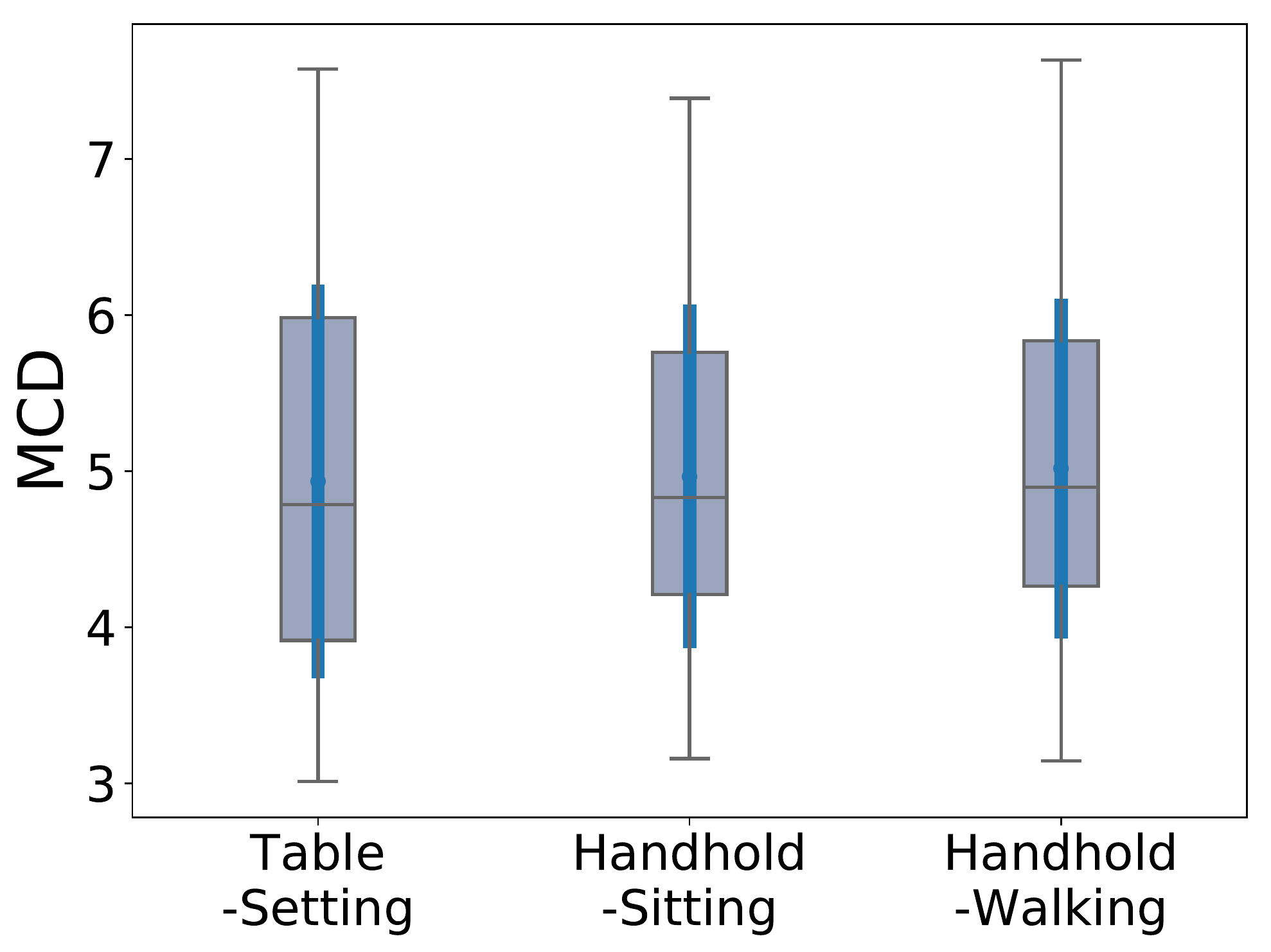}
\label{fig:pose}
}
\subfigure[Scenarios]
{\includegraphics[width=0.23\textwidth]{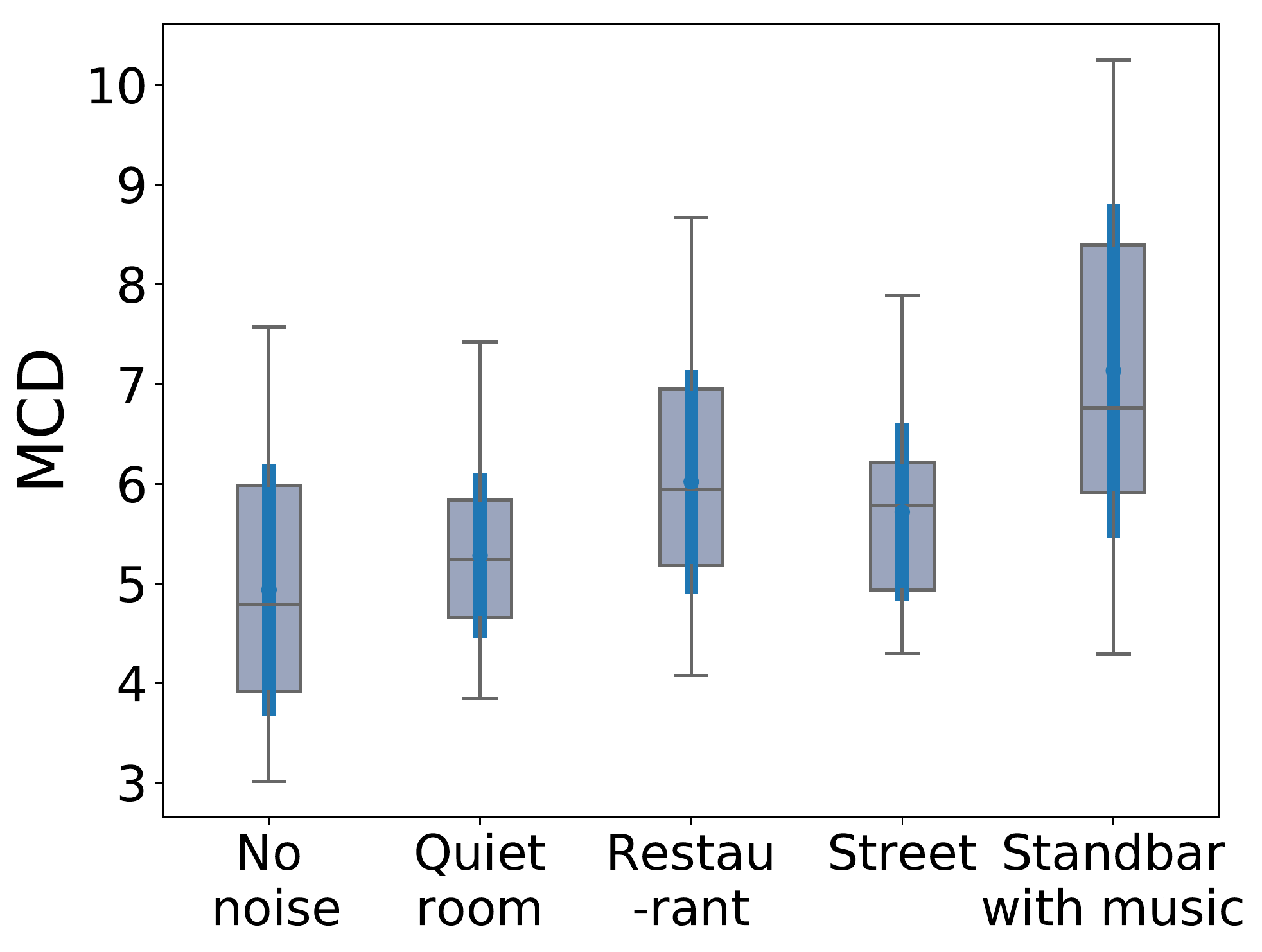}
\label{fig:scene}
}
\subfigure[Sampling rate]
{\includegraphics[width=0.23\textwidth]{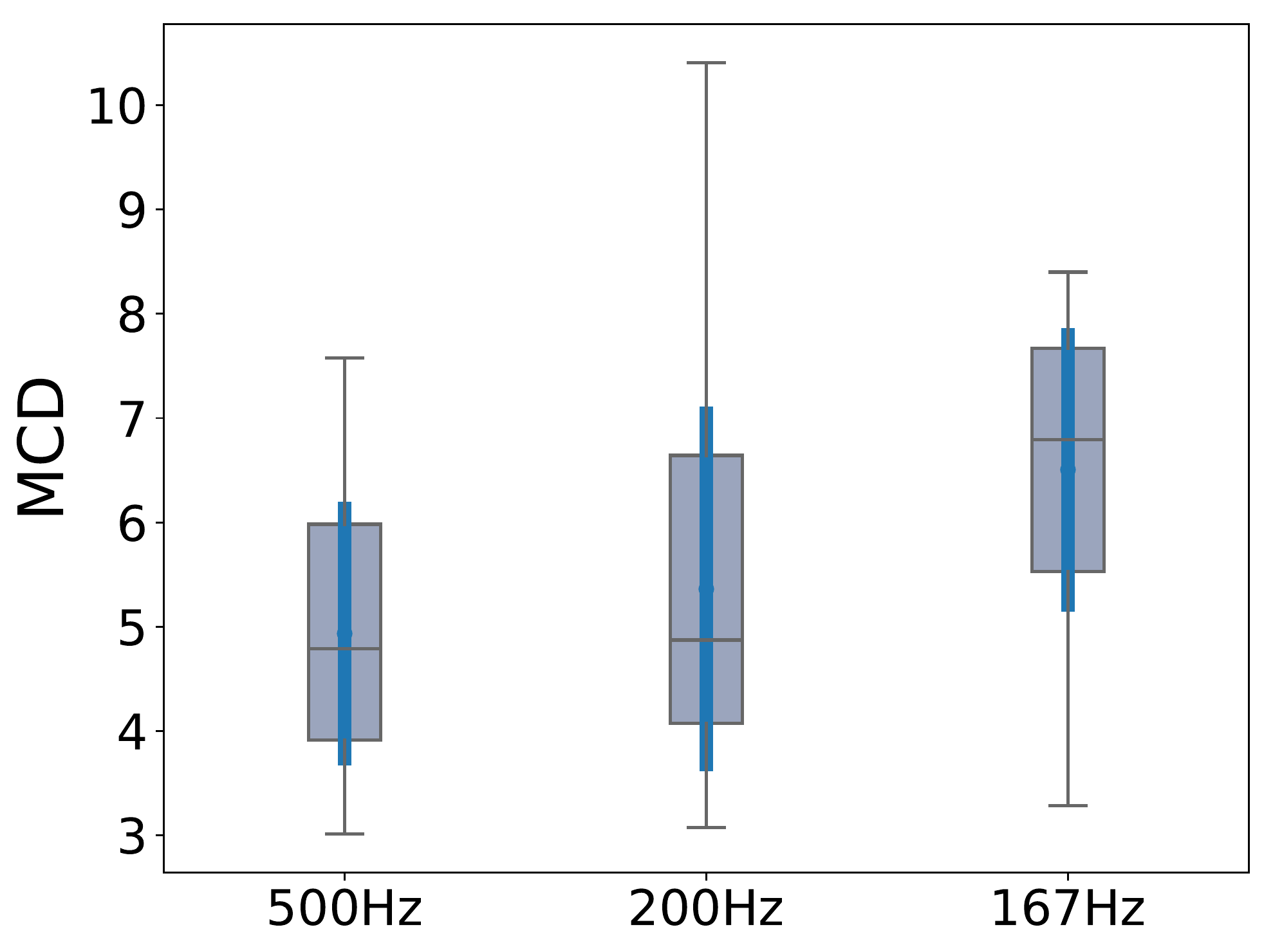}
\label{fig:sample}
}
\caption{Audio reconstruction performance with different settings}
\label{fig:various_setting}
\end{figure*}

\begin{figure}[t]
\centering
\subfigure[Original audio]
{\includegraphics[width=0.49\columnwidth]{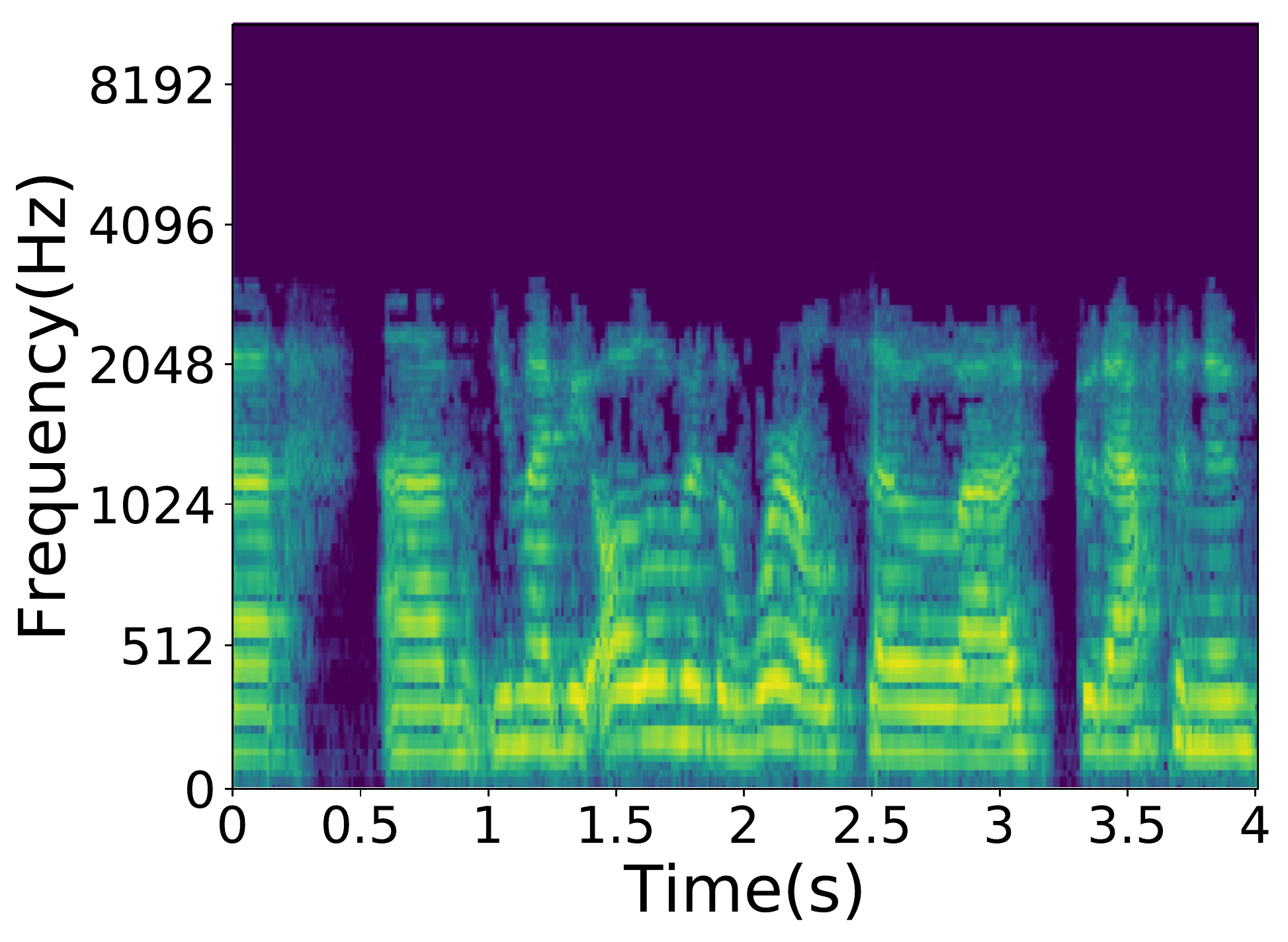}}
\subfigure[Reconstructed audio]
{\hspace{-12pt}\includegraphics[width=0.49\columnwidth]{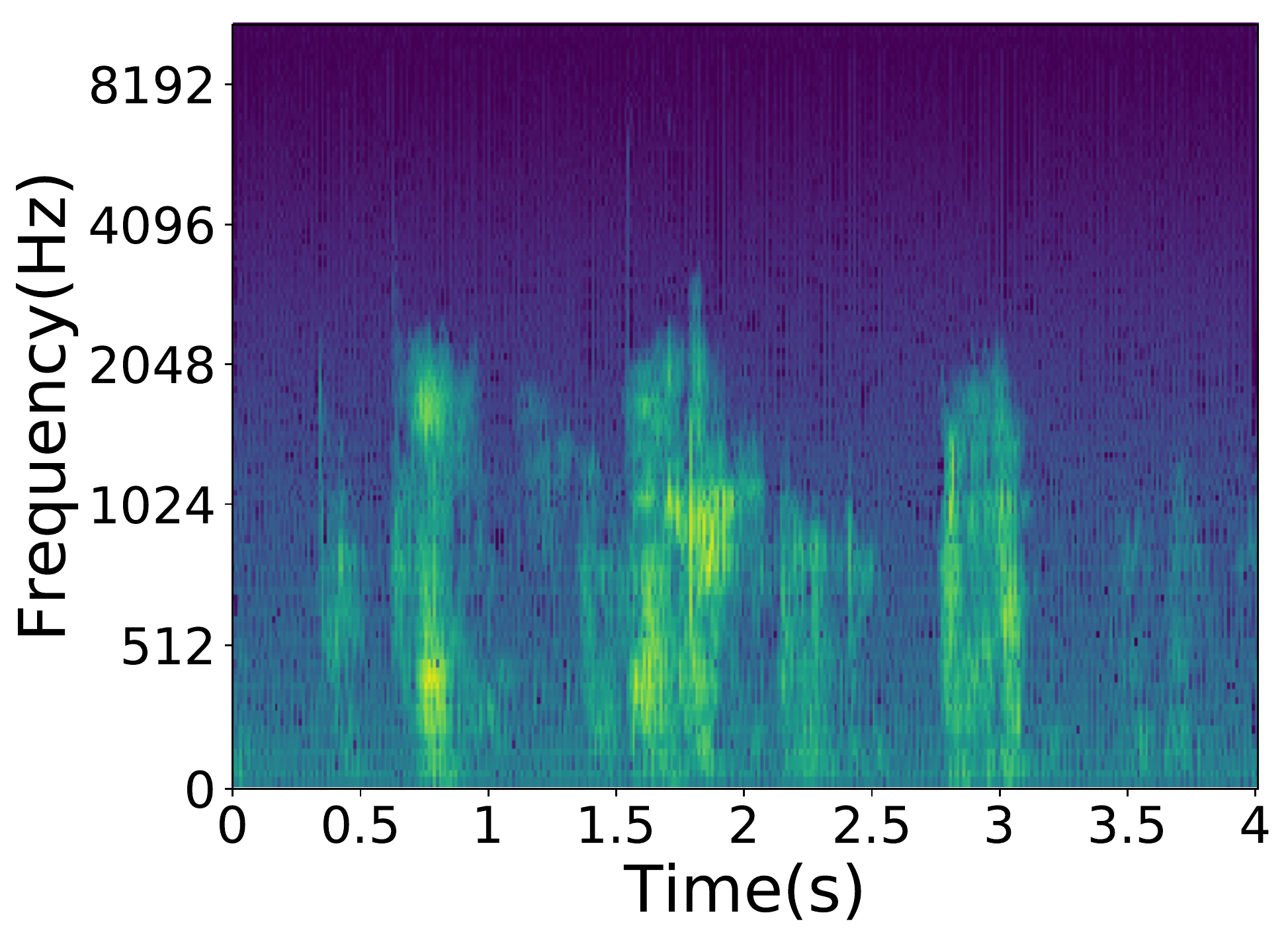}}
\caption{Mel spectrogram of original and reconstructed audio at 167Hz}
\label{fig:167Hz}
\end{figure}


\subsubsection*{\textbf{Impact of Volume}} 
Given that a user can play the sounds under different volumes, 
we collect the accelerometer data when the speaker plays the audio under different volume and test them on the model trained with the maximum volume. The performance of recovered audio at various volumes is shown in Fig.~\ref{fig:volume}, we observe that the MCD will increase with the volume decreases. This is because the vibration caused by the loudspeaker will weaken as the volume decreases, so the captured accelerometer data will diminish. As shown in Fig.~\ref{fig:volume}, we observe that most MCD is below 8, so we can reconstruct the audio through accelerometer data under these volume.

\begin{figure}
    \centering
    \includegraphics[scale=0.18]{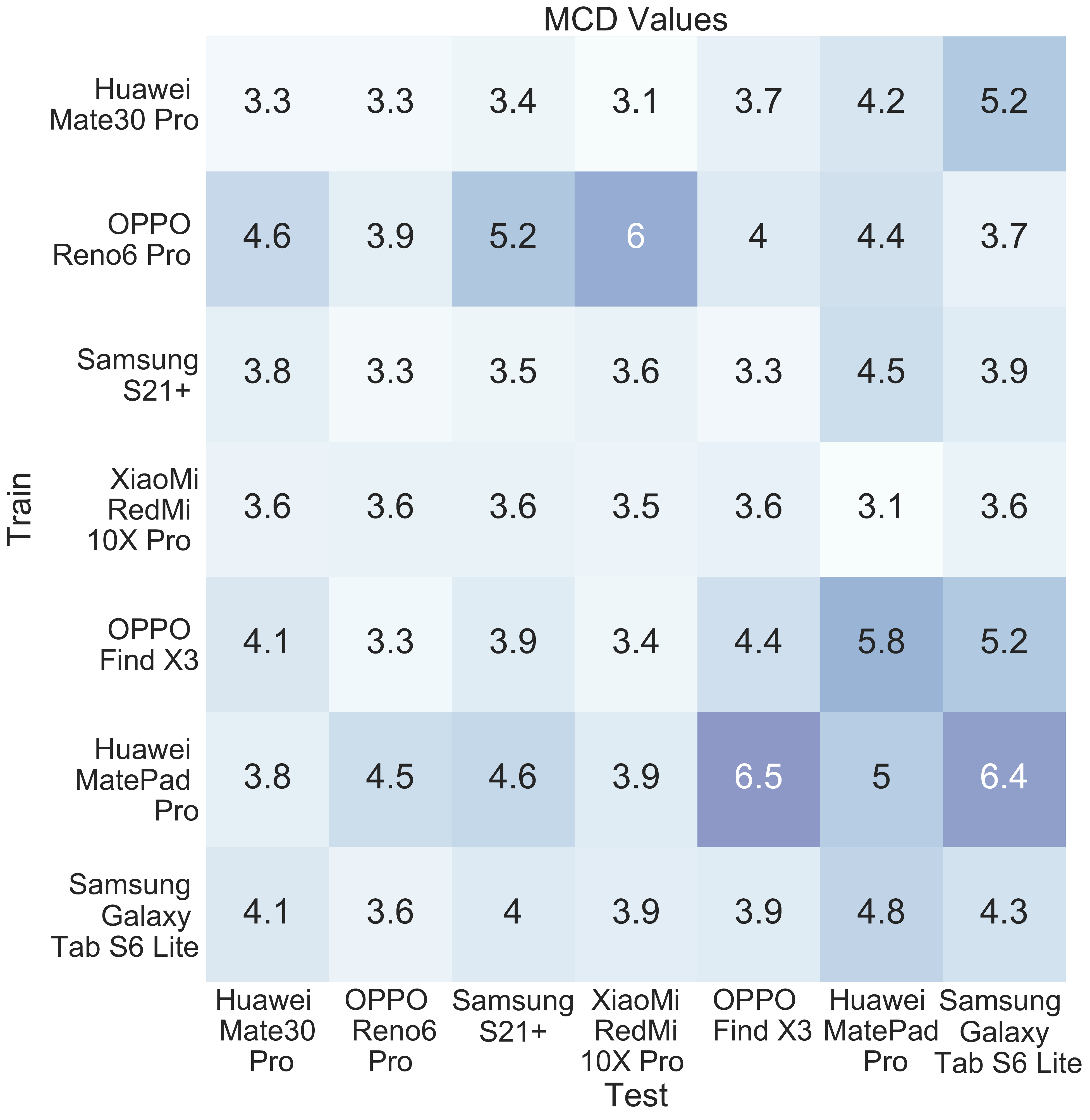}
    \caption{Generalizability of model on different mobile devices}
    \label{fig:phone_cross}
\end{figure}

\subsubsection*{\textbf{Impact of Phone Model}} 

The accelerometer sensor of each distinct mobile device can differ in terms of sampling rate and position on the motherboard. This can affect the quality of accelerometer data produced by the vibrations of a built-in speaker, which may also affect the generalizability of our cGAN model. 
To address this concern, we collect the accelerometer data
from five additional smartphones (Huawei Mate30 Pro, OPPO Reno6 Pro, Samsung S21+, OPPO Find X3, and XiaoMi RedMi 10X Pro) and two tablets (Huawei MatePad Pro and Samsung Galaxy Tab S6 Lite). According to the data in \cite{phonemarket}, the mobile phone brands we use account for $51.38\%$ mobile market share worldwide.
We train the model for each mobile phone and tablet. And for each model, we use the accelerometer data collected from other devices as the testing set to evaluate the generalizability of the model on other mobile phones. The MCD of reconstructed audio is shown in Fig.~\ref{fig:phone_cross}. Based on the results of the distinct smartphones, we observe that most MCD values are around $3$, and only few MCD values exceed $6$. 
Furthermore, we also test the generalizability between smartphones and tablets. The results in Fig.~\ref{fig:phone_cross} show not only that our attack works on tablets, but also that most of our models can generalize well across different phones and tablets. 

\begin{figure}[h]
\centering
\subfigure[Without High-pass filter.]
{\includegraphics[width=0.85\columnwidth]{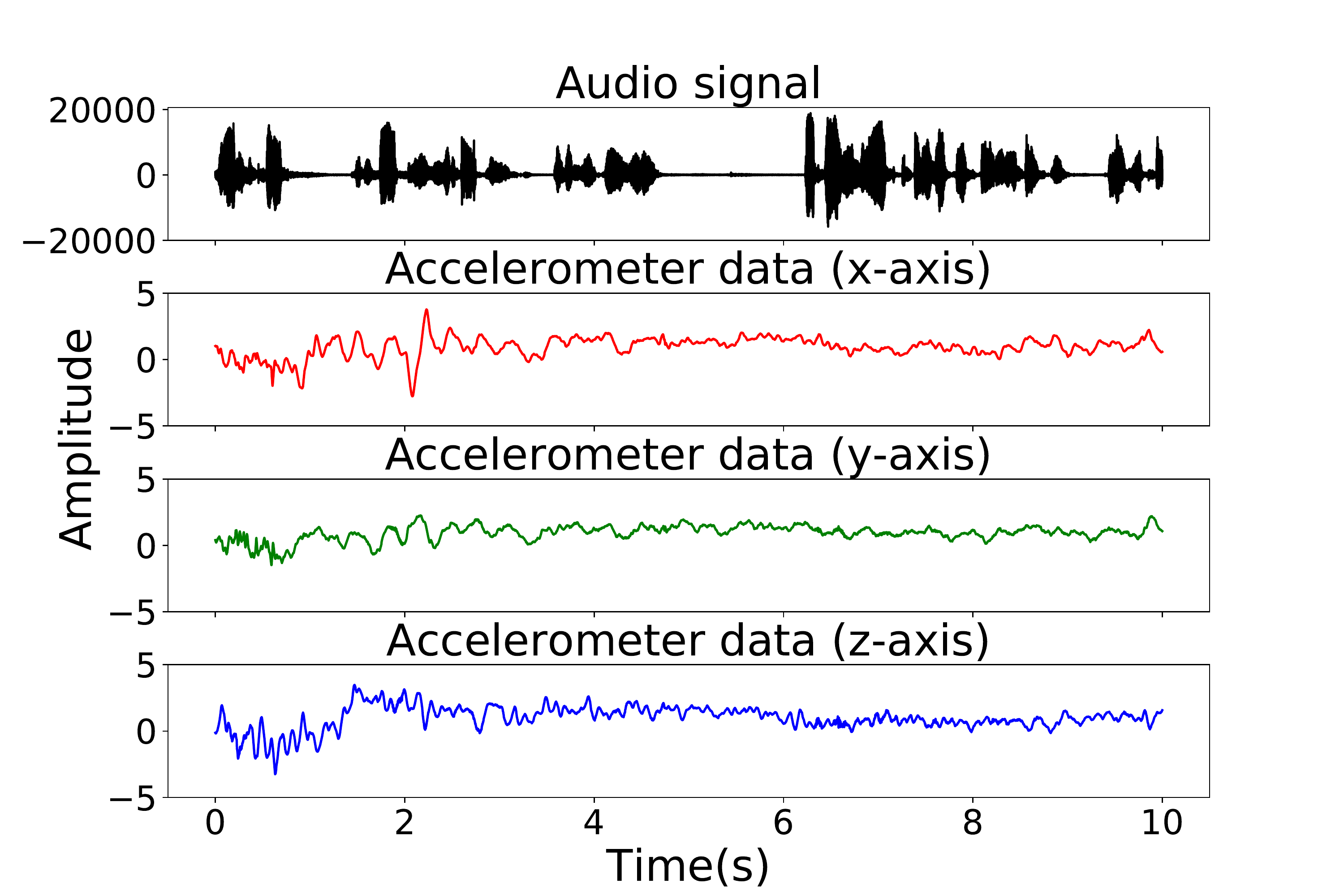}}
\hspace{-5pt}
\subfigure[With High-pass filter.]
{\includegraphics[width=0.85\columnwidth]{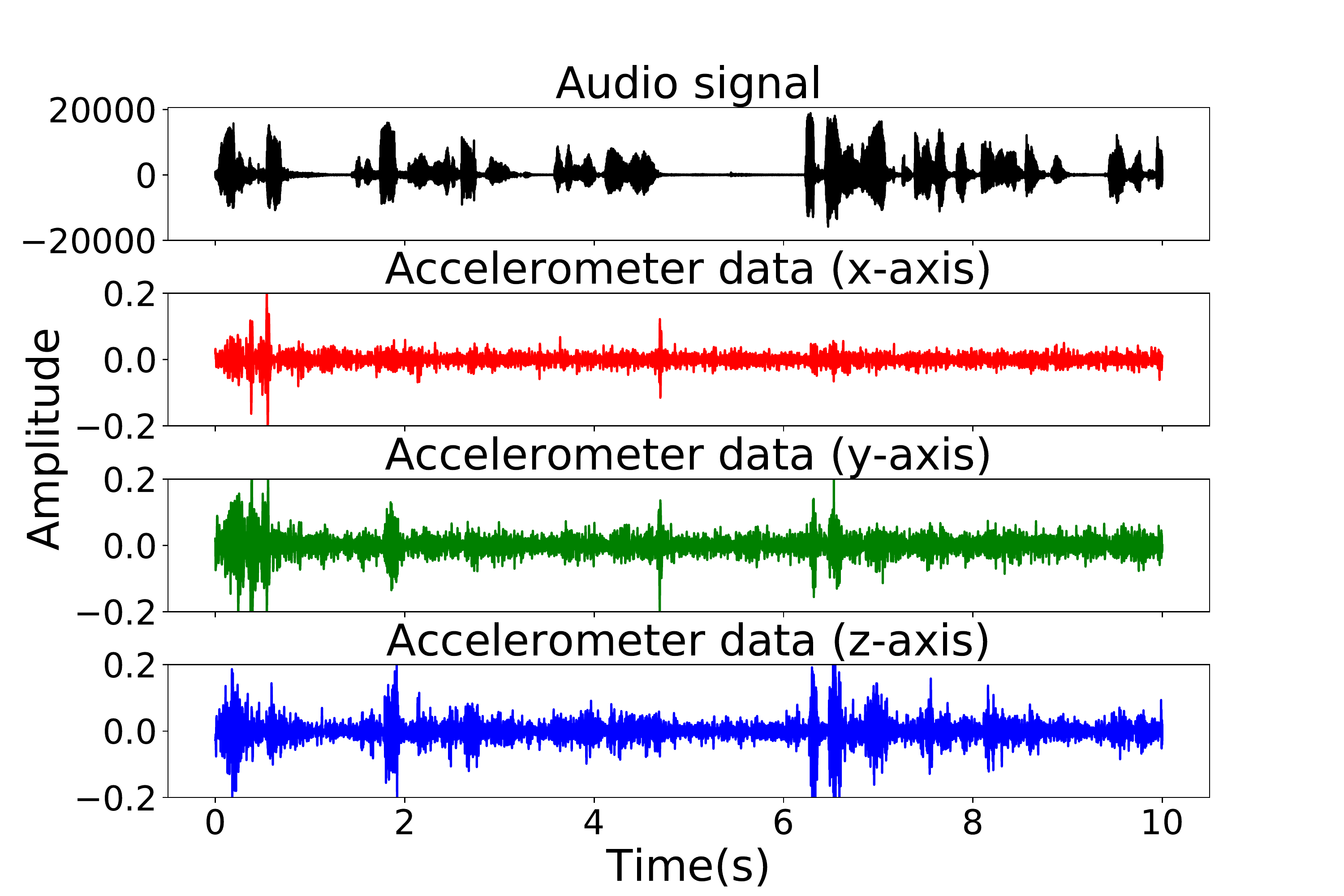}}
\caption{Accelerometer data when playing the audio while the user is walking: the high-pass filtering can effectively remove the movement related noise while preserving the audio-related vibrations.}
\label{fig:walking}
\end{figure}

\begin{figure}[t]
\centering
\subfigure[English speaker.]
{\includegraphics[width=0.48\columnwidth]{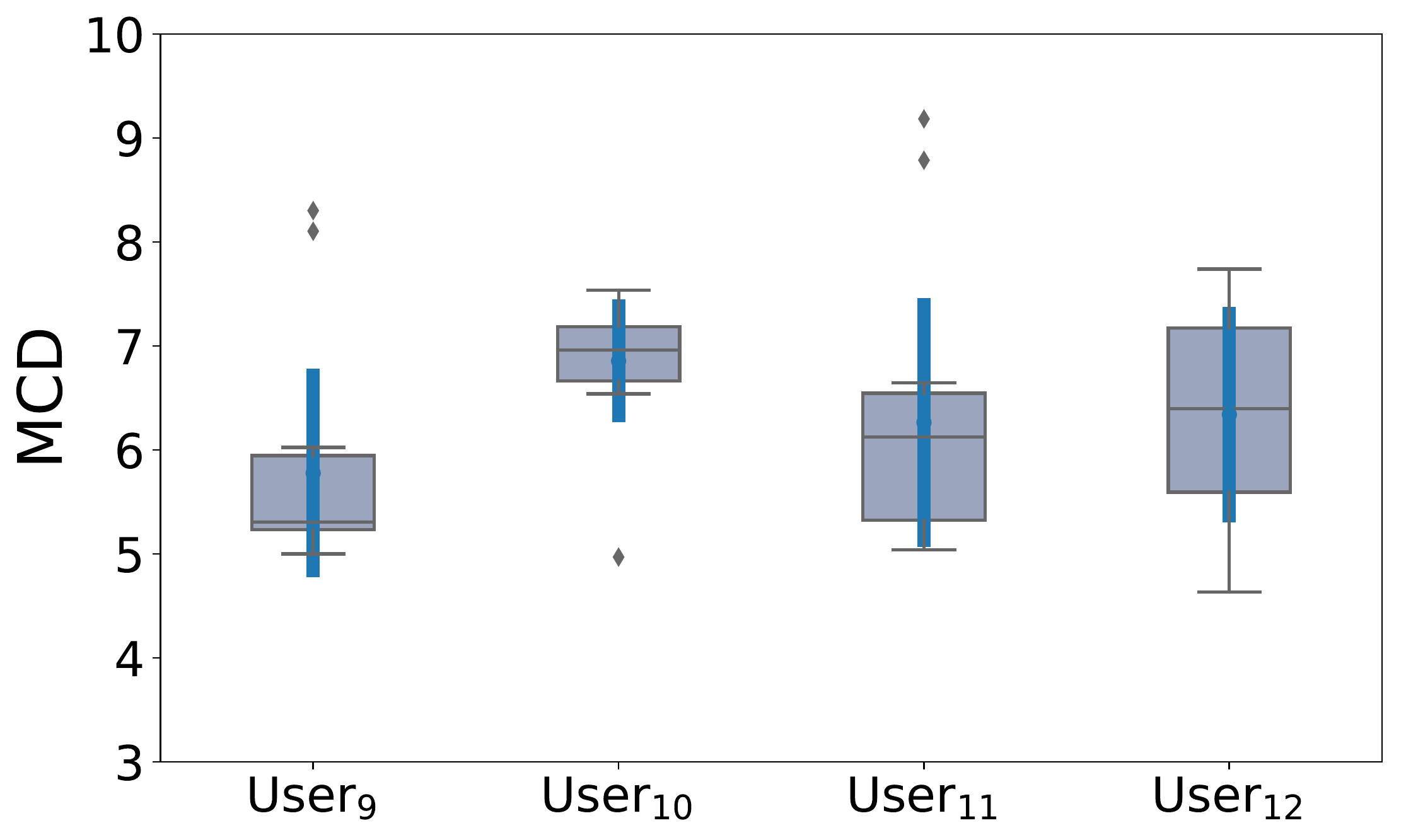}
\label{fig:English}}
\hspace{-5pt}
\subfigure[Chinese speaker.]
{\includegraphics[width=0.48\columnwidth]{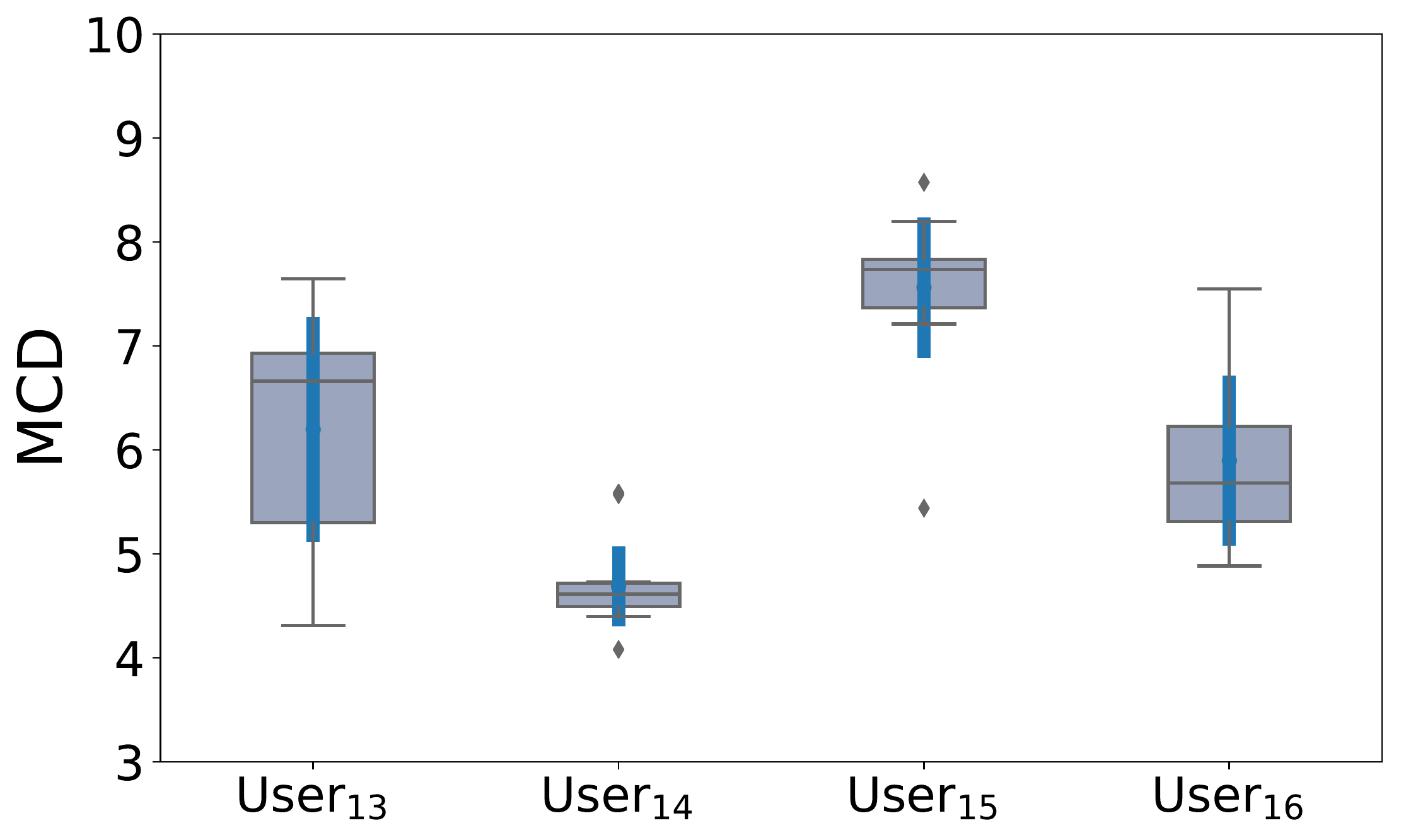}
\label{fig:Chinese}}
\caption{Audio reconstruction performance with different languages.}
\label{fig:mcd_language}
\end{figure}

\begin{figure*}[t]
\centering
\subfigure[English speaking female.]
{\includegraphics[width=0.45\columnwidth]{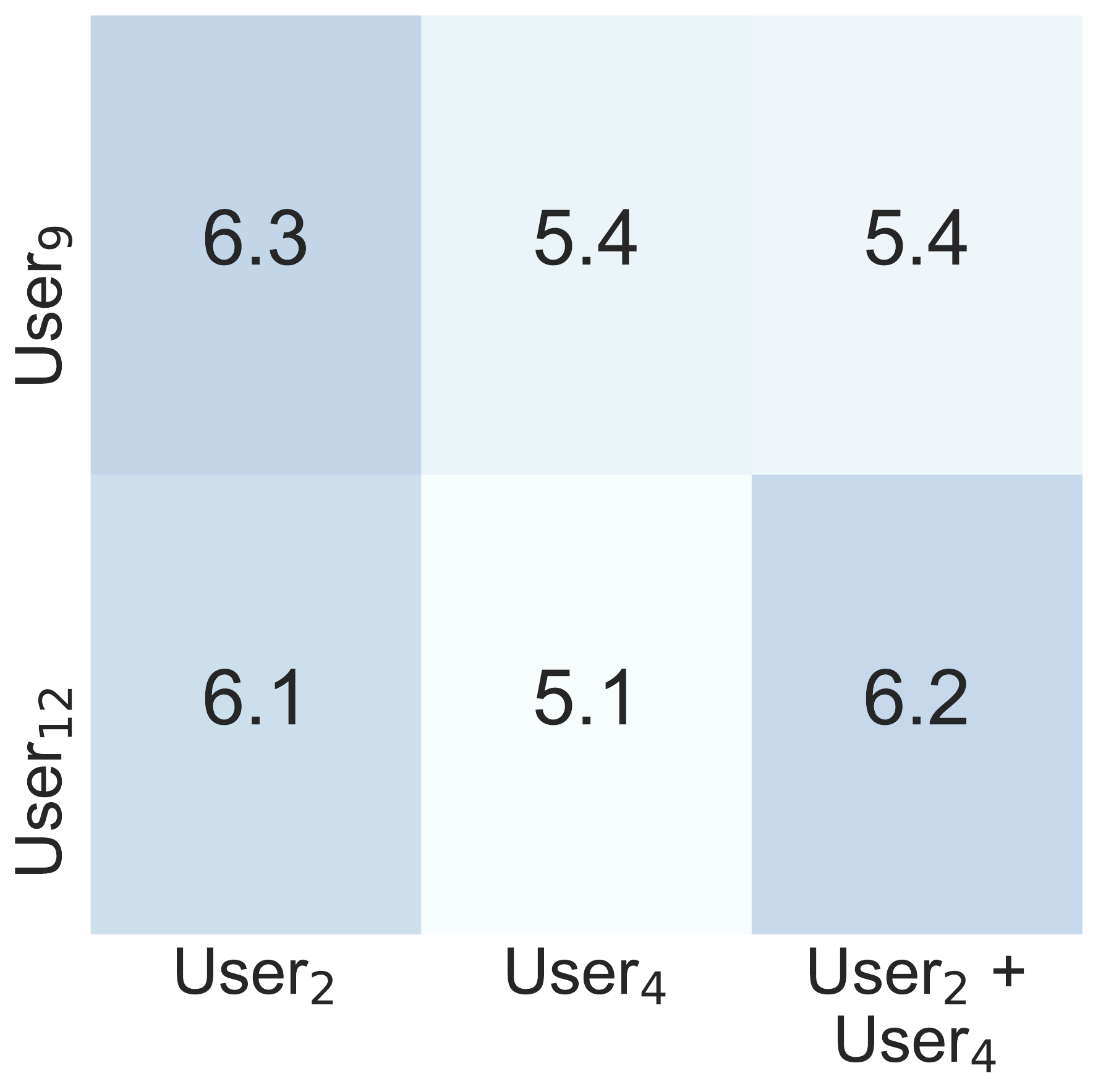}
\label{fig:eng_fe}}
\subfigure[Chinese speaking female.]
{\includegraphics[width=0.45\columnwidth]{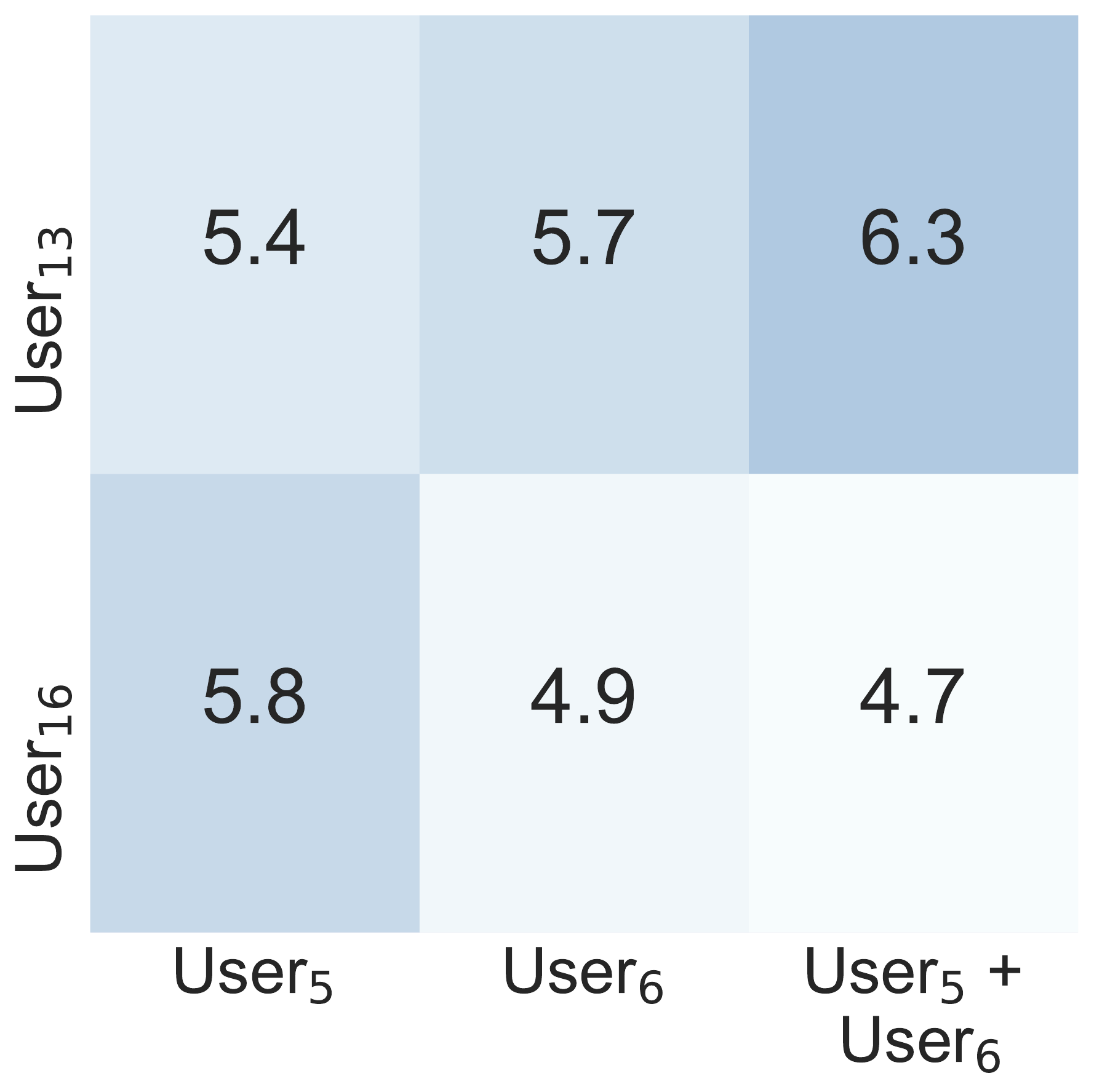}
\label{fig:chin_fe}}
\subfigure[English speaking male.]
{\includegraphics[width=0.45\columnwidth]{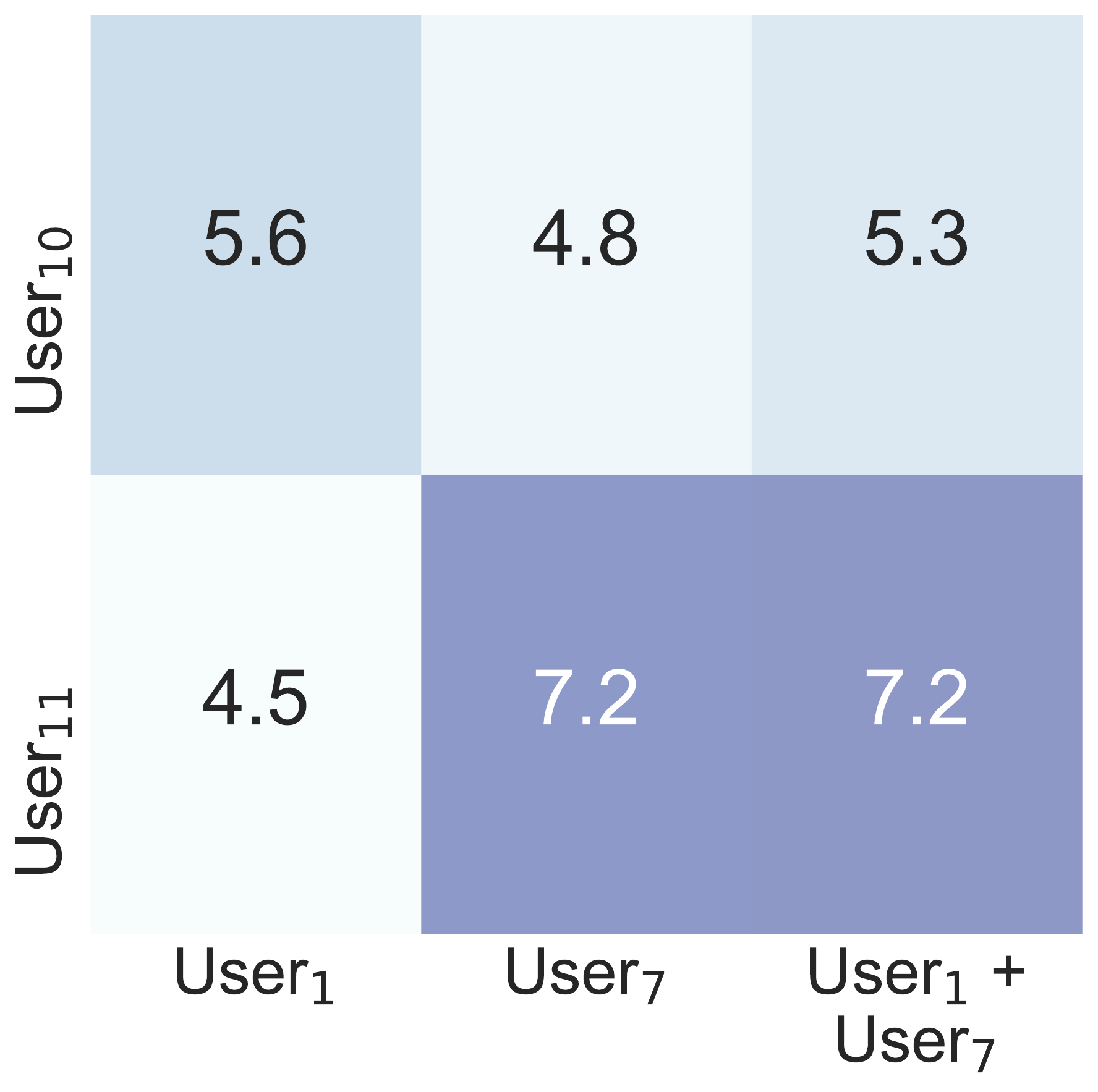}
\label{fig:eng_male}}
\subfigure[Chinese speaking male.]
{\includegraphics[width=0.45\columnwidth]{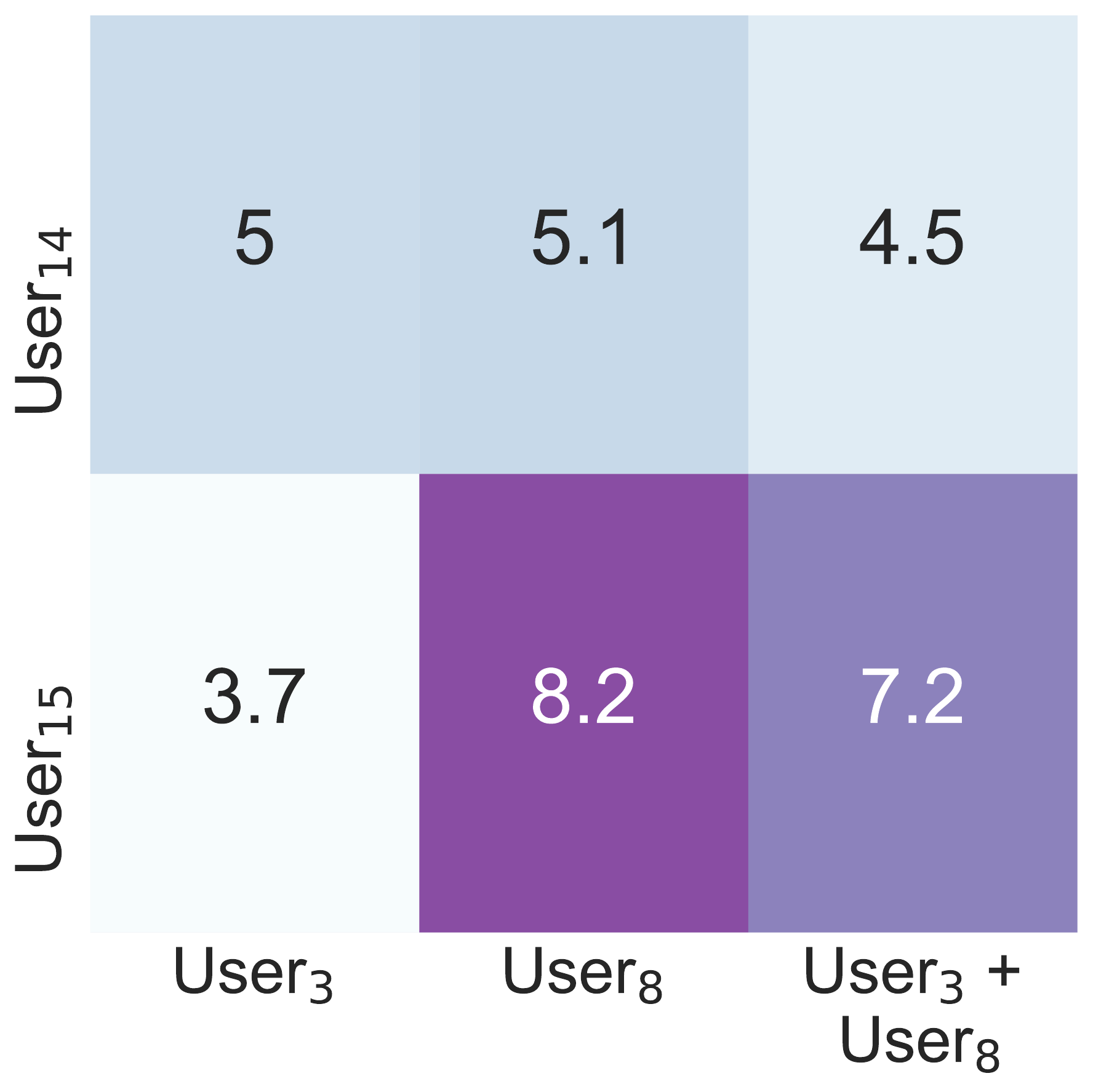}
\label{fig:chin_male}}
\caption{Performance of model generalization with cross-user training.}
\label{fig:heatmap}
\end{figure*}



\subsubsection*{\textbf{Impact of Placement}} 
To evaluate the impact of placement, the accelerometer data is collected when the smartphone was placed on a desk, held by a user who was sitting and walking respectively.
We believe the three types of positions represent the most common scenarios. 
We test these positions on the model trained with the phone placed on the desk. 
Since the placement of the device while the user holding the phone or walking affects the accelerometer data, it is a challenge to extract the voice-related accelerometer data in presence of noise related to human movement. 
To address this challenge, we apply a $20$Hz high-pass filter to remove the movement influence. 
Prior work \cite{zhang2015accelword} has shown that user movement such as walking and sitting is primarily concentrated in the lower frequency below $20$Hz. This means that the high-pass filter of $20$Hz would enable us to extract the voice-related vibrations from the noisy and mobility-influenced signal.  
Fig.~\ref{fig:walking} shows how the high-pass filter removes the movement noise while preserving the voice-related vibrations. 
Fig.~\ref{fig:pose} shows the MCD values under different conditions. 
We can observe that the high-pass filter clearly reduces the influence of movement and our model can achieve a similar MCD as the stationary case. 


\subsubsection*{\textbf{Impact of Scenario}}
During a video or voice call, 
the environmental sounds around the remote caller can influence the performance of our attack.
In this evaluation, we consider four common scenarios: no noise, a quiet room, a restaurant, a street with high pedestrian traffic, and standby with the music. 
To emulate these scenarios, we add their specific noises into the original audio. 
The results of audio reconstruction under different scenarios are shown in Fig.~\ref{fig:scene}. We can observe that most of the MCD value is lower except the scenario with music. The reason for the inferior performance of the music scenario is similar to the aforementioned performance of User$_6$. The blended audio signal has a wide spectrum range which somehow misleads our cGAN model.

\subsubsection*{\textbf{Impact of Sampling Rate}} To evaluate the influence of sampling rate on \pname, we collect the accelerometer data at the sampling rate of $167$Hz, $200$Hz, and $500$Hz for User$_1$ through User$_8$.
We reconstruct the audio based on the model trained with the sampling rate of $500$Hz. The performance of recovered audio at different sampling rates is shown in Fig.~\ref{fig:sample}.
As we expected, 
the MCD increases as the sampling rate decreases. 
We compare the Mel spectrogram of original audio and reconstructed audio under the sampling rate of $167$Hz in Fig.~\ref{fig:167Hz}, and the result demonstrates that our model can reconstruct partial information even at a sampling rate of only $167$Hz.

\subsubsection*{\textbf{Impact of Language}}
In this section, we train an English speaker model and a Chinese speaker model to validate the impact of languages. The English speaker model is trained on the data of User$_1$, User$_ 2$, User$_4$, and User$_7$, and its testing set is comprised of the data from User$_9$ to User$_{12}$. 
The MCD value of each testing user is shown in Fig.~\ref{fig:English}, we can observe that the mean value of MCD for each testing User is below $8$. 
The Chinese speaker model is trained by the data of User$_3$, User$_5$, User$_6$, and User$_8$, and its testing set is comprised of the data from User$_{13}$ to User$_{16}$. 
The MCD value of each testing user is shown in Fig.~\ref{fig:Chinese}, the mean values of them are also below $8$. 
This demonstrates that \pname works well in terms of different languages.


\subsubsection*{\textbf{Impact of Different User}}

As the training data could not include every user's speech samples (which have distinguishing features), it is necessary to reconstruct the audio of unknown users. 
To verify the generalization ability of \pname, we train three models using the data of User$_2$, User$_4$, and the data of User$_2$ and User$_4$ combined, and test on the data of User$_9$ and User$_{12}$. Note that they are all English-speaking females. As shown in Fig.~\ref{fig:eng_fe},

We can notice that the MCD in the case of unknown user is still below $8$. This demonstrates that our individual user model could reconstruct the speech of unknown users. 
We repeat the same experiments where the users are Chinese speaking females, English speaking males, and Chinese speaking males, the results are reported in Fig.~\ref{fig:chin_fe}, Fig.~\ref{fig:eng_male}, and Fig.~\ref{fig:chin_male}, respectively. 

We can also observe that when the model is trained using multiple users' data, the reconstruction performance could be worse than that of the model trained only using single user's data. 
This could be the fact that the diversity of speech has been introduced. Thus, training data with more users might not always help in reconstructing the audio of unknown users.
\begin{figure}[t]
    \centering
    \includegraphics[scale=0.32]{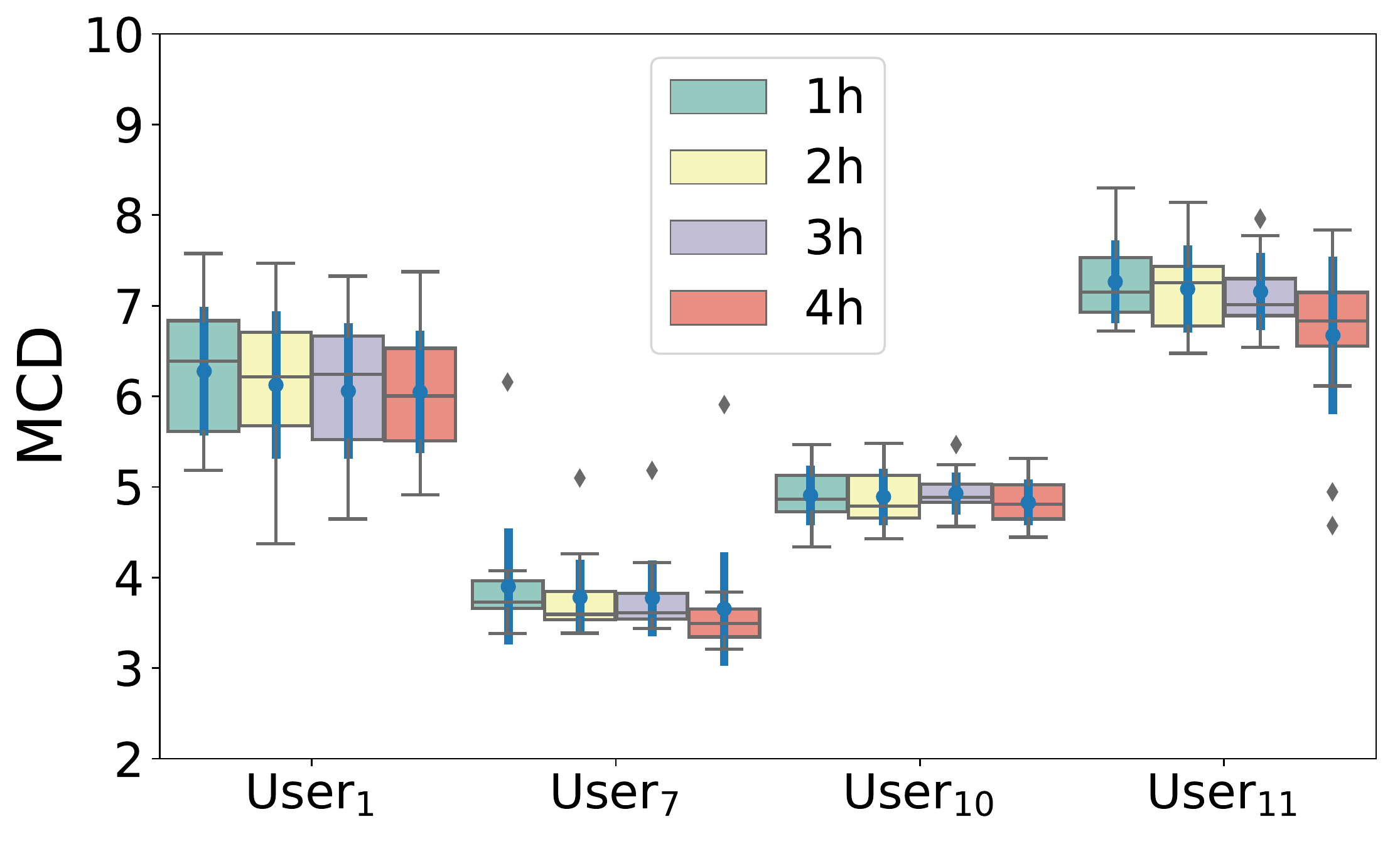}
    \caption{MCD under the various sizes of the training set}
    \label{fig:timelength}
\vspace{-10pt}
\end{figure}
Based on the above results, we further investigate the dataset size in terms of the length of time necessary to train a model that can effectively reconstruct other users' voices. 
In this experiment, we select the speech of User$_7$ (English speaker) as training sets and vary the datasets by one, two, three, and four hours. 
Then, we evaluate the model on the testing data of English-speaking User$_1$, User$_7$, User$_{10}$, and User$_{11}$. 
Fig.~\ref{fig:timelength} depicts the MCD values of the reconstructed audio. 
We can observe that almost all of MCD values of testing are lower than 8, which demonstrates that the models trained on 1$\sim$4 hours long datasets can effectively reconstruct the audio. 
In addition, we can notice that the performance of the model improves slightly along with the size of the dataset. A larger dataset will involve more training effort.
Hence, we need to reach a trade-off between the performance of audio reconstruction and training overhead. 


\section{Discussion}
\label{sec:discussion}

In this section, we discuss meaningful insights, possible countermeasures against our eavesdropping attack, the feasibility of other variants of GAN, limitations of cGAN, and future research directions.


In our experiments, we acquire the accelerometer data at the maximum sampling rate possible by using the \textit{SENSOR\_DELAY\_FASTEST} option but such a sampling rate depends on the smartphone manufacturer and the constraint of the operating system\cite{anand2018speechless}. 
For example, the Huawei Mate 40 Pro and Oppo Reno 6 Pro achieve a maximum sampling rate of $500$Hz and $420$Hz, respectively. 
At these sampling rates, \pname can effectively recover the speech information via accelerometer data.
However, Google recently proposed the new sampling rate limitation for motion sensors from Android 12 due to the exploit of such sensors for side channels attacks\cite{ratelimit}. 
According to this new security policy, an application needs to explicitly request user permission whether it accesses a motion sensor with a sampling rate higher than $200$Hz. 
However, in Section~\ref{sec:evalution} we test the effectiveness of our attack with the sampling rate of $167$Hz, $200$Hz, and $500$Hz. 
The experimental results in Fig.~\ref{fig:sample} show that \pname can still partially recover original audio even with a sampling rate of $167$Hz and $200$Hz. 

A possible countermeasure against our attack is to significantly decrease the maximum sampling rate of motion sensors for apps without the related user permission.
The \textit{SENSOR\_DELAY\_GAME} option (corresponding to a sampling rate of $50$Hz) already meets most requirements for the recognition of most human activities, which frequencies are below $30$Hz~\cite{qi2013adasense}. 
At this sampling rate, the effectiveness of our attack is pretty low since the accelerometer data can barely capture the unique features of different phonemes.
Therefore, the new security policy of Android 12 should require the user's permission when an application requests a sampling rate of accelerometer above $50$Hz rather than the current limit at $200$Hz.
Unfortunately, since updating a mobile operating system has minimum hardware requirements, many smartphones would run out-to-date operating systems thus they would still be vulnerable to our \pname attack.

 
Our \pname system has an unconstrained vocabulary since it learns the mapping between the accelerometer data and the Mel spectrogram for each phoneme pronunciation. 
Hence, the data in the training set needs to 
cover a sufficient number of different phonemes to achieve solid performance. 
To assess this, we can define the phoneme coverage as the ratio of the number of different phonemes covered by our training data to the total number of phonemes. 
For example, as the total number of phonemes in the English language is $48$, an audio sample that contains $24$ different phonemes has a phoneme coverage of $0.5$. In our experiments, even if the audio samples contain thousands of words, we cannot ensure (despite very likely) that they have a full phoneme coverage (i.e., $1.0$).
We will also consider the variations for the same phonemes in the phoneme coverage computation and further investigate their impact on the audio reconstruction, as pointed out in Fig.~\ref{fig:variation_high}. 
In future work, we will investigate suitable methods to automatically calculate the phoneme coverage of audio samples for a better training dataset.

GAN has been extensively studied for its strong data generation ability. Among the many variants of GAN in literature, we adopt cGAN to perform the conversion from accelerometer data to the corresponding audio. 
Such a variant is particularly suitable for this task for two reasons: 1) cGAN accepts an input condition to control the output;
2) cGAN can realize the one-to-one mapping which allows the generator to learn the mapping between conditions and outputs. 
Unfortunately, the other variants of GAN either do not accept an input condition (such as DCGAN\cite{radford2016unsupervised}, EBGAN\cite{zhao2016energy}, LSGAN\cite{mao2017least}, WGAN\cite{arjovsky2017wasserstein}, etc.) or they do not achieve a one-to-one mapping between inputs and outputs (such as CycleGAN\cite{zhu2017unpaired}, StyleGAN\cite{karras2019style}, etc).
To the best of our knowledge, cGAN is the only variant that fits the requirements of our task.

Our cGAN-based approach also has some limitations.
In particular, the inputs and outputs of our cGAN are image-like two-dimensional data.
Hence, we have to transform the accelerometer data to spectrogram by using SFTF and then transform the output Mel spectrogram to an audio waveform by using the Griffin-Lim algorithm. 
However, such transformation leads to the information loss of the signal phase, which may distort the reconstructed audio. 
Furthermore, cGAN is known for its difficulty of training in terms of model tuning and computation overhead.
In future work, we plan to explore possible neural network-based approaches which can directly process the time series data to avoid the information loss caused by the spectrogram conversion.

\section{CONCLUSION}
\label{sec:conclusion}
In this paper, we propose an accelerometer eavesdropping system \pname that reconstructs the audio played by the built-in speaker from accelerometer data. With \pname, an adversary can reconstruct unconstrained words from accelerometer data, so it can be extensively used in voice and video calls, voice navigation, voice assistant, and other scenarios. We implement and extensively evaluate \pname on different smartphones and users, achieving high accuracy under various settings and scenarios. 

\section*{Acknowledgement}
We would like to thank the anonymous reviewers and
the shepherd for their insightful comments. This work is supported by the National Key Research and Development Program of China (Grant No. 2021YFB3100400) and National Natural Science Foundation of China (Grant No. 61832012).



\appendix
\subsection{Detailed parameters of mobile devices}
\label{appendix:parametersmodels}
We list the detailed information on the different models of the mobile devices in Table~\ref{tab:para}. We can notice that while the maximum accelerometer sampling rate is within a range 416$\sim$500Hz on smartphones, it is significantly lower on tablets (i.e., 200$\sim$250Hz). Despite this difference, our attack scheme shows consistency among multiple devices.
\begin{table}[ht]
\centering
\scalebox{0.925}{
\begin{tabular}{lcccc}
\hline
                           \textbf{Model}         & \textbf{Type} & \textbf{System Version} & \textbf{\begin{tabular}[c]{@{}c@{}}Screen\\ Size \end{tabular}} & \textbf{\begin{tabular}[c]{@{}c@{}}Acceler.\\ MSR \end{tabular}} \\ \hline
\textbf{Huawei Mate40 Pro}          & Phone         & HarmonyOS 2.0         & 6.76 in.             & 500Hz                                    \\ \hline
\textbf{Huawei Mate30 Pro}          & Phone         & HarmonyOS 2.0         & 6.53 in.             & 500Hz                                    \\ \hline
\textbf{OPPO Reno6 Pro}             & Phone         & Android 11              & 6.55 in.             & 420Hz                                    \\ \hline
\textbf{SamSung S21+}               & Phone         & Android 11              & 6.70 in.             & 416Hz                                    \\ \hline
\textbf{\begin{tabular}[l]{@{}l@{}} XiaoMi RedMi\\ 10X Pro \end{tabular}}       & Phone         & Android 11              & 6.57 in.             & 418Hz                                    \\ \hline
\textbf{OPPO Find X3}               & Phone         & Android 11              & 6.70 in.             & 425Hz                                    \\ \hline
\textbf{Huawei MatePad Pro}         & Tablet        & HarmonyOS 2.0         & 10.80 in.            & 250Hz                                    \\ \hline
\textbf{\begin{tabular}[l]{@{}l@{}}Samsung Galaxy\\ Tab S6 Lite \end{tabular}} & Tablet        & Android 11              & 10.40 in.            & 200Hz                                    \\ \hline
\end{tabular}
}
\caption{Detailed properties of different mobile devices (Acceler. MSR stands for Accelerometer Maximum Sampling Rate)}
\label{tab:para}
\end{table}

\subsection{Relationship between MCD and reconstruction performance at word level}
\label{appendix:mcdrecontruct}
We further explore the relationship between MCD and reconstruction performance at the word level and randomly select some sample results to present in Table~\ref{tab:mcd}.
We believe that even if the model cannot reconstruct all the words in a sentence, we can infer the missing words from the context. Besides, we can also resort to recent Natural Language Processing (NLP) techniques (such as BERT \cite{devlin2018bert}) to infer the semantics of sentences even with missing words.

\begin{table}[ht]
\centering
\begin{threeparttable}
\begin{tabular}{ p{0.5cm}p{3.5cm}p{3.5cm}}
\hline
  \textbf{MCD}  &  \textbf{Original Audio}&  \textbf{Reconstructed Audio} \\ \hline \hline
  2$\sim$3 & Zheng ji bi sai, jiang jin shi wan, mei & Zheng ji bi sai, jiang jin shi wan, mei\\ \hline 
  3$\sim$4 & you might be all over the world so good afternoon &  *   are be all over the world so good afternoon\\ \hline
  4$\sim$5 & We had a barrel like this down in our basement, filled with cans of food and water & We had a barrel like this *  in our basement, filled with cans of food and * \\ \hline
  5$\sim$6& This is our product line. We have a very clean product line, we think we have the best notebooks in the business & This is our product line. *  most *  clean product line, we think we have the best  *   in the business   \\ \hline
  6$\sim$7& Talking about here at Ted is that ther're right in the middle of rainforest was of some solar panels the & Talking about here at Ted is * * *  in *  * the middle of rainforest was of some solar panels the \\ \hline
  7$\sim$8 & Community could have light for I think it was about half an hour each evening and there is the chief in all his&  Community could have light * .  *  think *  *  *  half * hour each evening and there is the chief in all his\\ \hline
\end{tabular}
\vspace{5pt}
    \caption{MCD and corresponding reconstructed results (* represents the word we cannot recognize).}
    \label{tab:mcd}
\end{threeparttable}
\end{table}

\subsection{The effect of coverage of dataset diversities on model performance}
\label{appendix:coverageeffect}
The diversity of a person's speech mainly lies in two aspects: speed and frequency. People's speech speed can often change depending on the speaker's mood and context. Using the data with normal speech speed for training and a much faster or slower speed for testing will lead to an unsatisfactory result of speech reconstruction. The pronunciation frequency of people in different emotional states can also be different. For example, the pronunciation frequency can be lower when the mood is low and relatively higher when the mood is excited. Therefore, the change of frequency is also within our consideration.

\begin{figure}[t]
\centering
\subfigure[Speed]
{\includegraphics[width=0.4\textwidth]{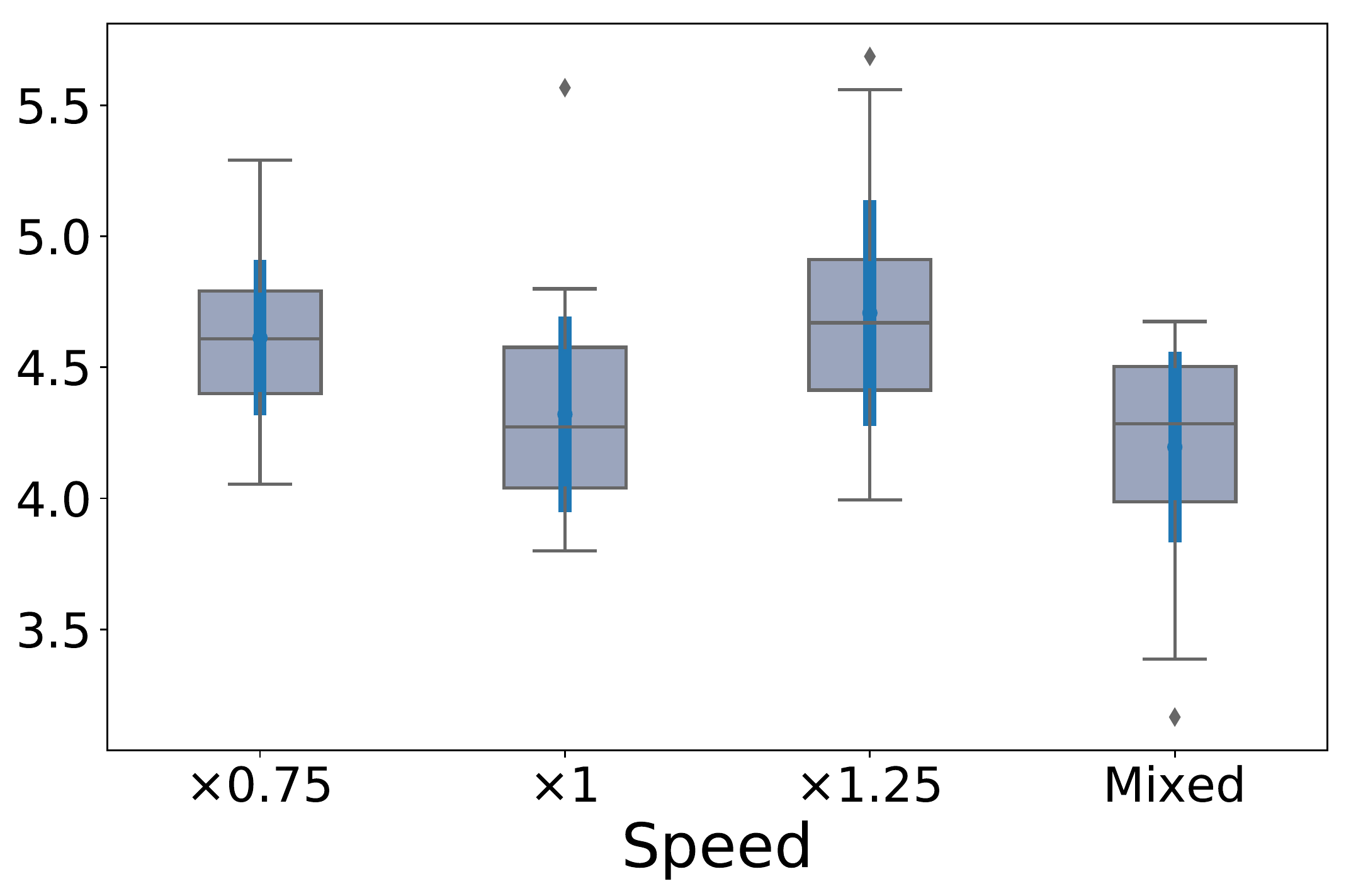}
\label{fig:speed}
}\\
\subfigure[Frequency]
{\includegraphics[width=0.4\textwidth]{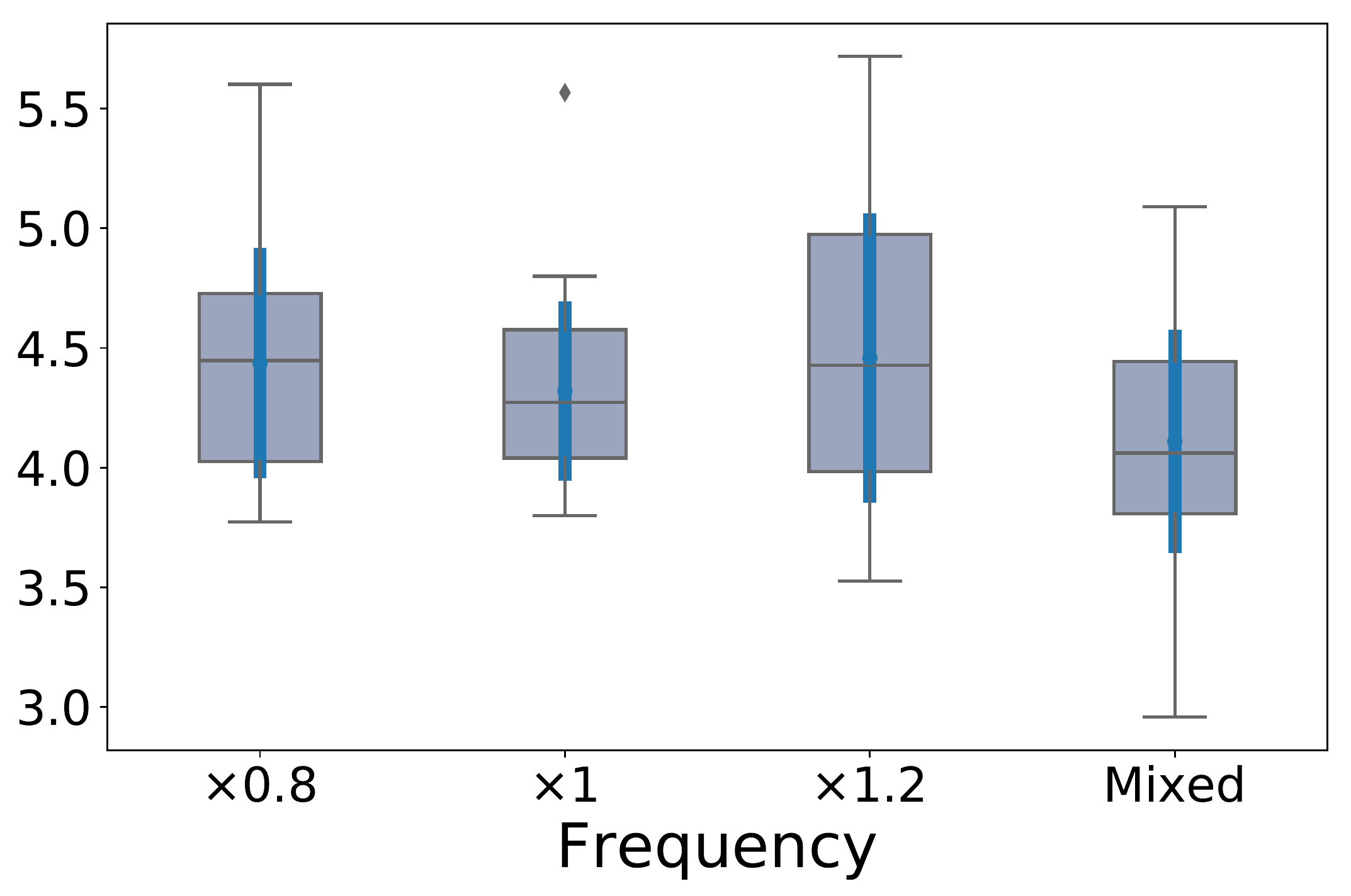}
\label{fig:frequency}
}\\
\subfigure[All]
{\includegraphics[width=0.4\textwidth]{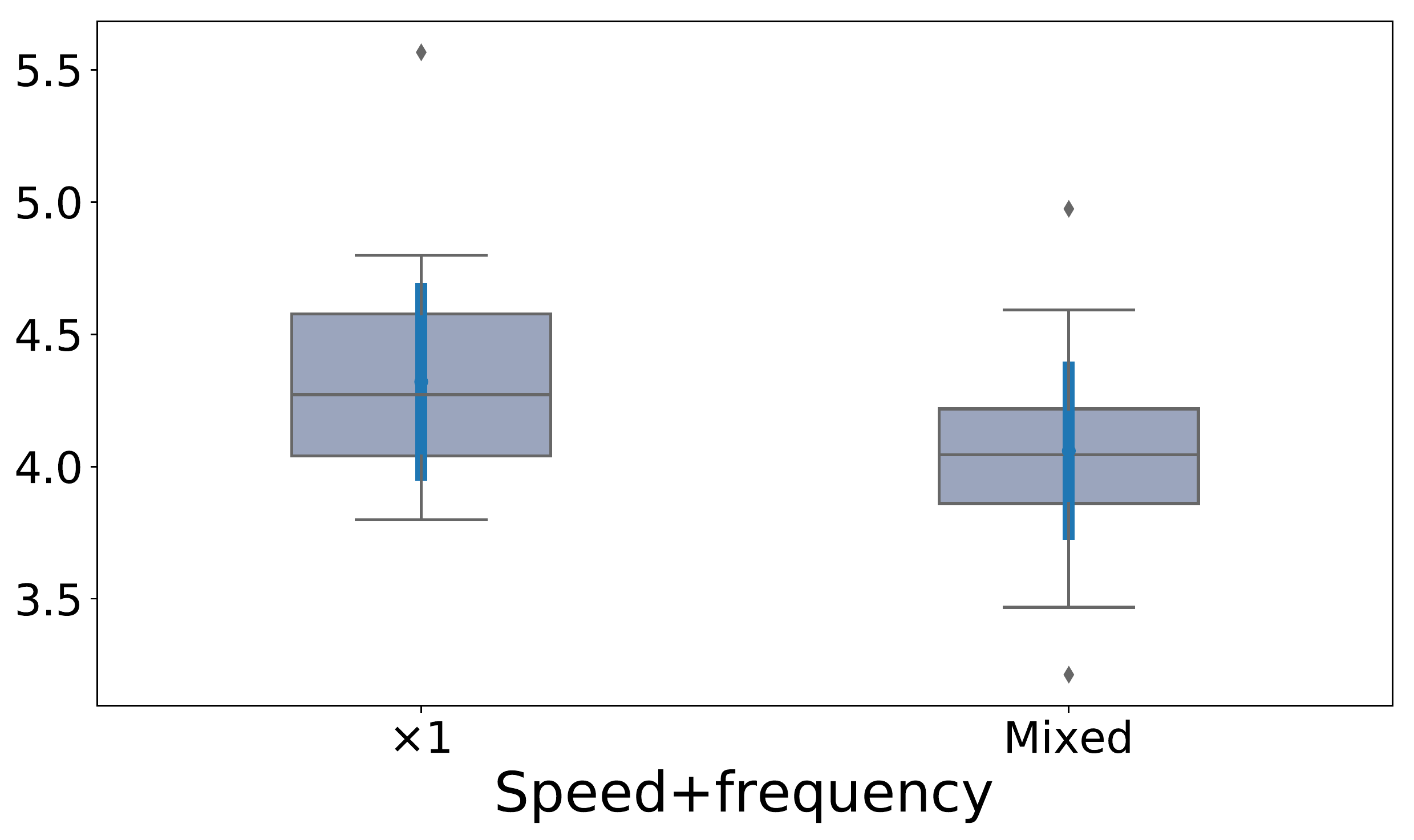}
\label{fig:all}
}
\caption{Audio reconstruction performance with speed and frequency diversity.}
\label{fig:diversity}
\end{figure}

We perform a systematic evaluation of the diversity. We prepare the following datasets:
\begin{enumerate}
    \item For the speed,
    \begin{enumerate}
        \item $\times 0.75$
        \item $\times 1.0$
        \item $\times 1.25$
        \item mixed dataset including $\times 0.75$, $\times 1.0$, and $\times 1.25$
    \end{enumerate}
    \item For the frequency,
    \begin{enumerate}
        \item $\times 0.8$
        \item $\times 1.0$
        \item $\times 1.2$
        \item mixed dataset including $\times 0.8$, $\times 1.0$, and $\times 1.2$
    \end{enumerate}
    \item mixed dataset including 1.d and 2.d
\end{enumerate}
As shown in Fig.~\ref{fig:diversity}, the model with more diversity achieves a better performance in general. In terms of speed, the model (1.d) based on mixed datasets achieves 2.9$\%$ improvement than single dataset as shown in Fig.~\ref{fig:speed}. In terms of frequency, the model (2.d) based on mixed datasets achieves 4.9$\%$ improvement as shown in Fig.~\ref{fig:frequency}. The model (3) with most diversity achieves the largest improvement of 6.0$\%$ as shown in Fig.~\ref{fig:all}. 

\subsection{The transferability between different users}
\label{appendix:confusion matrix}

\begin{figure}[t]
    \centering
    \includegraphics[width=0.45\textwidth]{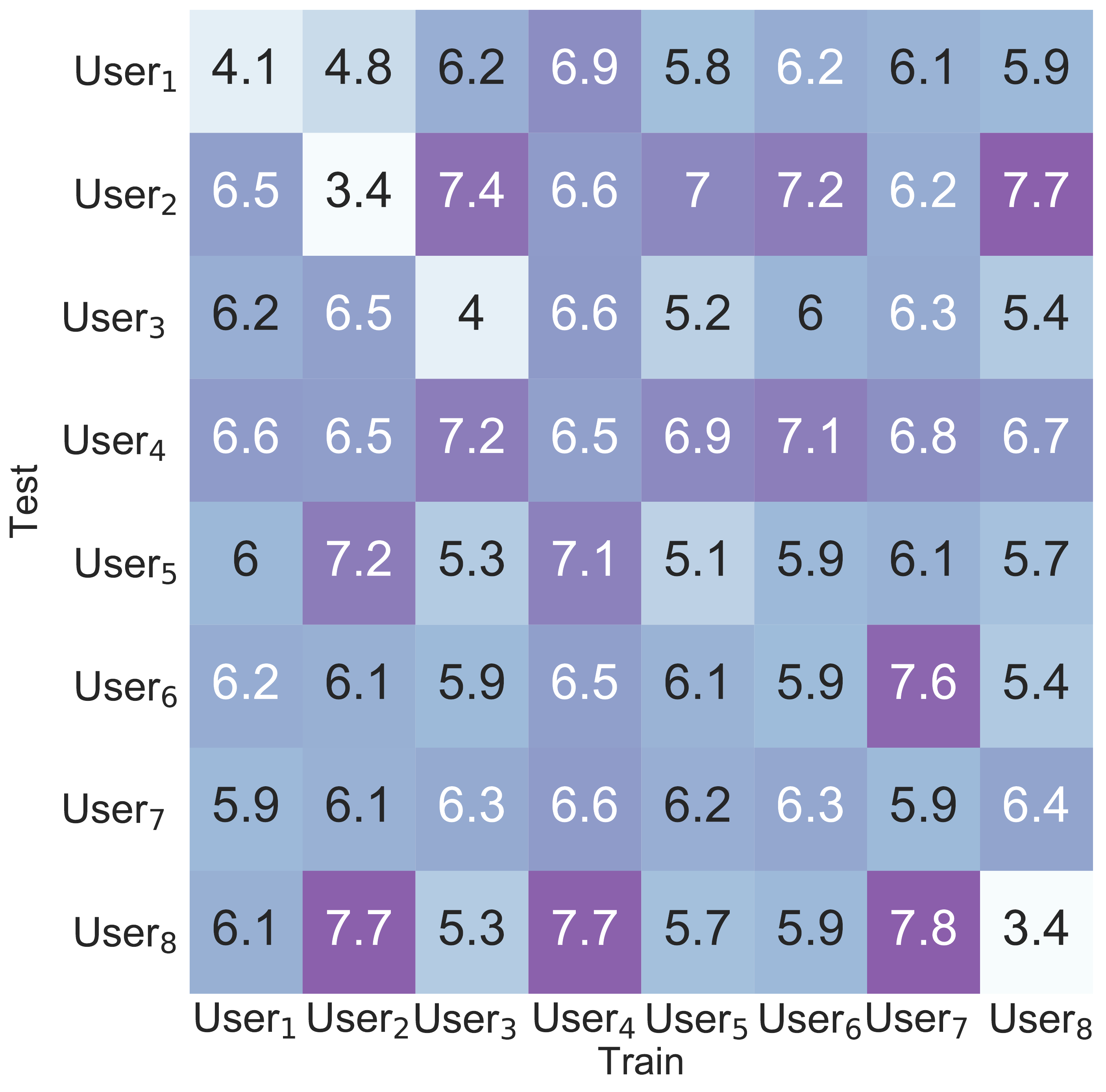} 
    \caption{Performance of model generalization with cross-user training}
    \label{fig:confusion}
\end{figure}

We test the transferability between all users, as shown in Fig.~\ref{fig:confusion}. The results show our model can generalize well under cross-user training.

\end{document}